\documentclass[journal]{IEEEtran}
\usepackage{cite}
\ifCLASSINFOpdf
  \usepackage[dvipdfmx]{graphicx} 
  \usepackage{bmpsize}
  \graphicspath{{../pdf/}{../jpeg/}}
  \DeclareGraphicsExtensions{.pdf,.jpeg,.png}
\else
  \usepackage[dvipdfmx]{graphicx} 
  \usepackage{bmpsize}
  \graphicspath{{../eps/}}
  \DeclareGraphicsExtensions{.eps}
\fi
\usepackage{amsmath}
\usepackage{algorithm}
\usepackage{algorithmic}
\usepackage{array}

\usepackage{stfloats}
\usepackage{multirow}
\usepackage[normalem]{ulem}
\useunder{\uline}{\ul}{}
\usepackage{subcaption}
\usepackage[justification=centering]{caption}
\usepackage{epstopdf}

\usepackage{float}
\usepackage{amsmath}
\usepackage{color}
\usepackage{chemformula}
\usepackage[version=4]{mhchem}
\usepackage[table,xcdraw]{colortbl}
\newcommand\copyrighttext{%
	\footnotesize \textcopyright 2023 IEEE. Personal use of this material is permitted. Permission from IEEE must be obtained for all other uses, in any current or future media, including reprinting/republishing this material for advertising or promotional purposes, creating new collective works, for resale or redistribution to servers or lists, or reuse of any copyrighted component of this work in other works.}
\newcommand\copyrightnotice{%
	\begin{tikzpicture}[remember picture,overlay]
		\node[anchor=south,yshift=10pt] at (current page.south) {\fbox{\parbox{\dimexpr\textwidth-\fboxsep-\fboxrule\relax}{\copyrighttext}}};
	\end{tikzpicture}%
}

\begin{document}
\title{Electrochemical Parameter Identification for Lithium-ion Battery Sources in Self-Sustained Transportation Energy Systems} % 或者说 for practical implementation

\author{Yuxuan~Gu,~\IEEEmembership{Student Member,~IEEE,}
        Jianxiao~Wang,~\IEEEmembership{Member,~IEEE,}
        Yuanbo~Chen,~\IEEEmembership{Student Member,~IEEE,}
        Kedi~Zheng,~\IEEEmembership{Member,~IEEE,}
        Zhongwei~Deng,~\IEEEmembership{Member,~IEEE,}
        and~Qixin~Chen,~\IEEEmembership{Senior Member,~IEEE}% <-this % stops a space
\thanks{Manuscript submitted Dec. 08, 2022; revised Jan. 25, 2023; and accepted Mar. 10, 2023. This work was supported by the International (NSFC-NWO) Joint Research Project of National Natural Science Foundation of China under Grant 52161135201 and the Major Smart Grid Joint Project of the National Natural Science Foundation of China and State Grid under Grant U2066205.}
\thanks{Y.~Gu, Y.~Chen, K.~Zheng and Q.~Chen are with the Department of Electrical Engineering, Tsinghua University, 100084, Beijing, China.}
\thanks{J.~Wang is with the National Engineering Laboratory for Big Data Analysis and Applications, Peking University, 100871, Beijing, China.}
\thanks{Z.~Deng is with the College of Mechanical and Vehicle Engineering, Chongqing University, 400044, Chongqing, China.}
\thanks{Q.~Chen is the corresponding author (e-mail:qxchen@tsinghua.edu.cn).}
}


\markboth{IEEE Transactions On Industry Applications}%
{Y. Gu \MakeLowercase{\textit{et al.}}: Bare Demo of IEEEtran.cls for IEEE Journals}



%\IEEEoverridecommandlockouts
%\IEEEpubid{\makebox[\columnwidth]{978-1-5386-5541-2/18/\$31.00~\copyright2018 IEEE \hfill}
%	\hspace{\columnsep}\makebox[\columnwidth]{ }}
\maketitle
%\IEEEpubidadjcol
\copyrightnotice

\begin{abstract}
Lithium-ion battery (LIB) sources have played an essential role in self-sustained transportation energy systems and have been widely deployed in the last few years. To realize reliable battery maintenance, identifying its electrochemical parameters is necessary. However, the battery model contains many parameters while the measurable states are only the current and voltage, inducing the identification inherently an ill-conditioned problem. A parameter identification approach is proposed, including the experiment, model, and algorithm. Electrochemical parameters are first grouped manually based on the physical properties and assigned to two sequenced tests for identification. The two tests named the quasi-static test and the dynamic test, are compressed on time for practical implementation. Proper optimization models and a sensitivity-oriented stepwise (SSO) optimization algorithm are developed to search for the optimal parameters efficiently. Typically, the Sobol method is applied to conduct the sensitivity analysis. Based on the sensitivity indexes, the SSO algorithm can decouple the mixed impacts of different parameters during the identification. For validation, numerical experiments on a typical NCM811 battery at different life stages are conducted. The proposed approach saves about half the time finding the proper parameter value. The identification accuracy of crucial parameters related to battery degradation can exceed 95\%. Case study results indicate that the identified parameters can not only improve the accuracy of the battery model but also be used as the indicator of the battery SOH.
\end{abstract}


% \fi
\IEEEpeerreviewmaketitle

\section*{Nomenclature}
\addcontentsline{toc}{section}{Nomenclature}
\subsection*{Notations}

\begin{IEEEdescription}[\IEEEusemathlabelsep\IEEEsetlabelwidth{$k/K/\mathcal{K}$}]
	\item[$\kappa$] Ionic conductivity of the electrolyte (S/m).
	\item[$\sigma_s^{\pm}$] Electrical conductivity of active materials in the negative and positive electrodes (S/m).
	\item[$\sigma_f^{\pm}$] Electrical conductivity of SEI/CEI over active particles in negative/positive electrodes (S/m).
	\item[$\varepsilon_{s}^{\pm}$] Volume fraction of the active material in the negative and positive electrodes.
	\item[$\varepsilon_{e}^{\pm}$] Volume fraction of the electrolyte in the negative and positive electrodes and the separator.
	\item[$\delta_f^{\pm}$] SEI/CEI film thickness over active materials in the negative/positive electrodes (nm).
	\item[$\rho^{\pm}$] Density of active materials in the negative and positive electrodes (kg/m$^3$).
	\item[$A^{\pm,\mathrm{sep}}$] Projected area of the negative and positive electrodes and the separator (m$^2$).
	\item[$L^{\pm,\mathrm{sep}}$] Thickness of the negative and positive electrodes and the separator (m).
	\item[$c_s^{\pm}$] Bulk-averaged \ce{Li^+} concentration of active materials in the negative and positive electrodes (mol/m$^3$).
	\item[$c_{ss}^{\pm}$] Surface \ce{Li^+} concentration of active materials in the negative and positive electrodes (mol/m$^3$).
	\item[$c_e^{\pm}$] \ce{Li^+} concentration of the electrolyte in the negative and positive electrodes (mol/m$^3$).
	\item[$C_Q$] Battery capacity (mAh).
	\item[$D_e$] Diffusion coefficient of the electrolyte (m$^2$/s).
	\item[$D_s^{\pm}$] Diffusion coefficient of active materials in the negative and positive electrodes (m$^2$/s).
	\item[$I$] Applied current (A).
	\item[$k_r^{\pm}$] Reaction rate coefficient of reactions in the negative and positive electrodes (A$\cdot$m$^{2.5}$/mol$^{1.5}$).
	\item[$M^{\pm}$] Molecular mass of active materials in the negative and positive electrodes (kg/mol).
	\item[$R_c$] Contact resistance of the current collector ($\Omega$).
	\item[$R_f^{\pm}$] SEI/CEI film resistance over active materials in the negative/positive electrodes ($\Omega\cdot$m$^2$).
	\item[$R_r^{\pm}$] Active particle radius (m).
	\item[$t_0^+$] Ion transference number of the electrolyte.
	\item[$\Delta t$] Discrete time step (s).
	\item[$l$] Index of the time step.
	\item[$U_{\mathrm{OCP}}^{\pm}$] Equilibrium potential functions of reactions in the negative and positive electrodes (V).
	\item[$V$] Battery voltage (V).
\end{IEEEdescription}

\section{Introduction}
\IEEEPARstart{W}{ith} the rising concern of carbon emissions in the transportation sector, the self-sustained highway transportation energy system provides a promising way to achieve environmentally friendly and low carbon goals by exploiting abundant solar energy sources~\cite{ICPS54075}. To cope with the intermittency and volatility in solar power and increase energy conversion efficiency, the lithium-ion battery (LIB) technology is an essential option due to its advantages such as long lifespan, high energy density, and so on~\cite{3110938}. However, batteries suffer from continuous degradation stress when providing energy conversion services. They may even encounter severe conditions such as thermal runaways, indicating the necessity of the state of health (SOH) estimation. Typically, the lack of SOH information may increase safety risks, e.g., the LIB storage system explosion in dahongmen, Beijing, in 2021~\cite{jin_explosion_2021}, was due to the detonation of flammable gas in a few weak batteries operating at unreasonable points.

Generally, the SOH estimation of LIBs is achieved via model-driven approaches. Specifically, parameters in LIB models are identified based on the measurements timely. By tracking the changes in crucial parameters, the estimation of SOH can be realized~\cite{7420729}. The commonly used models include equivalent circuit models (ECMs) and electrochemical models (EMs). Due to their low complexity, ECMs are more prevalent in practice. However, identifying the parameters of the ECM mainly brings two drawbacks. First, the generalization ability of ECMs can hardly be ensured with limited measurement data since the operating characteristics of the battery can vary significantly under different working protocols, i.e., the reliability of ECM parameters is doubtful. Second, ECM parameters mainly serve to fit the output rather than representing the specific underlying mechanisms inside the battery, i.e., they lack concrete physical interpretations that can help better understand the degradation process of the battery.

Unlike ECM parameters, EM parameters have explicit physical meanings directly related to physical processes or chemical reactions inside the battery. Identifying these parameters can not only help us improve the modeling accuracy of the battery, but also serve as efficient indicators of the battery SOH. In addition, the development of simplified EMs in current studies made the identification of EM parameters achievable~\cite{GU2023126192}.

Considering the investment cost of battery sources in transportation energy systems, the commonly available measurements include the current and voltage. Such low-dimensional measurement data makes the identification approach challenging on both theoretical and practical levels. From the perspective of cybernetics, the battery can be viewed as a single-input (the current) and single-output (the voltage) system. However, the number of EM parameters is large, and the impacts of different parameters on the final output are mixed. They make the identification task inherently an ill-conditioned problem, i.e., directly identifying all parameters synchronously may yield a solution that fits the input-output well but is unreasonable and lacks physical significance. On the practical level, to increase the number of observable parameters as much as possible, the battery needs to be tested under various current profiles, making the identification process time-consuming and impractical for the operators of transportation energy systems. Meanwhile, the non-linearity of the EM further makes the search for optimal parameters consume more time and computation resources. 

Generally, the parameter identification process includes three stages, i) defining parameters to be identified, ii) designing the identification test, and iii) searching optimal parameter values. The literature review follows the order of the three stages above to demonstrate the achievements and shortages of current studies in each stage.

EMs usually contain more parameters compared to ECMs. Identifying all the parameters based on current-voltage measurements is not realistic. Thus, crucial or identifiable parameters are mainly considered. To decide these parameters, existing research can be categorized into two approaches, the manual approach and the numerical approach. In the manual approach, the parameters to be identified are decided by the researchers according to their knowledge or requirements. Ref.~\cite{gao_development_2022} chose to identify diffusion coefficients and reaction rate coefficients. Ref.~\cite{lyu_situ_2019} listed nine parameters for identification, including the volume fractions of active materials, diffusion coefficients, and so on. Ref.~\cite{li_simplified_2021} identified thermodynamics, slow-dynamics, and fast-dynamics parameters respectively. In the numerical approach, the parameters to be identified are determined via sensitivity analysis. Ref.~\cite{song_parameter_2018} applied the Fisher information matrix, and Cramer-Rao bound analysis to obtain the parameter identifiability. Ref.~\cite{bizeray_identifiability_2019} constructed the identifiable parameter set according to the maximum expected degrees of freedom of the battery model after partially non-dimensionalizing. To compute the sensitivity, the One-at-a-Time (OAT) method is applied in refs.~\cite{li_parameter_2020,li_data-driven_2022}. In refs.~\cite{hu_control_2020, bi_adaptive_2020}, the analytical expressions to the sensitivity of crucial parameters are derived.

After the parameters to be identified are decided, specific tests are designed to obtain the required measurements. Ref.~\cite{wang_fractional-order_2015} reviewed commonly-used standard tests, including the static capacity test, hybrid pulse power characterization (HPPC) test, and so on. Ref.~\cite{li_simplified_2021} adopted the HPPC test to identify fast-dynamics parameters and the pulse discharge (PD) test to identify slow-dynamics parameters. Ref.~\cite{namor_parameter_2017} designed the low C-rate galvanostatic test to identify parameters related to the electrode capacity, the PD test to identify the ohmic resistance and reaction rate coefficients, and the galvanostatic intermittent titration test (GITT) to identify diffusion coefficients. A similar design was proposed in ref.~\cite{li_aging_2020}; the main difference is this work used a high-frequency alternating current test to identify the ohmic resistance. Ref.~\cite{li_physics-based_2017} cut the entire measurement records into segments and assigned each parameter to the segment with the highest sensitivity index for identification. A series of PD tests with different current amplitudes and durations were designed in ref.~\cite{lyu_situ_2019}. In addition to common measurements, the electrochemical impedance spectroscopy (EIS)~\cite{chu_control-oriented_2020}, and the incremental capacity (IC) curve~\cite{yang_evaluation_2022,lin_evolution_2022} were included in the data source for identification.

After the measurements for identification are obtained, a proper optimization algorithm should be specified to search for the optimum. Existing algorithms can be divided into exact algorithms and heuristic algorithms. Exact algorithms determine the searching direction and forward step in a deterministic manner. Refs.~\cite{deng_implementation_2017,zheng_lithium-ion_2018} applied the non-linear least squares regression (Levenberg-Marquardt method). To avoid the gradient computation, ref.~\cite{xu_parameter_2020} adopted the pattern search non-linear optimization (PSNO). However, due to the high non-linearity of the EM, exact algorithms are likely to be trapped in the local optima. Unlike exact algorithms, heuristic algorithms decide each searching step in a stochastic manner and are efficient in finding the global optimum. Classic heuristic algorithms have been widely used in existing studies, e.g., the particle swarm optimization (PSO)~\cite{zhang_novel_2018,fan_systematic_2020,rahman_electrochemical_2016} and the genetic algorithm (GA)~\cite{xiong_online_2020,feng_co-estimation_2020,li_parameter_2016}. In addition, various novel algorithms were proposed as well, including the mixed-swarm-based cooperative PSO~\cite{xu_state_2020, hu_lithium-ion_2018}, the homotopy method~\cite{masoudi_parameter_2015}, the hybrid adaptive PSO and simulated annealing method~\cite{zhou_adaptive_2021}, the surrogate-assisted teaching-learning-based optimization~\cite{9263350}, the hybrid improved harmony search and Bayes neural network~\cite{kim_data-efficient_2019} and the cuckoo search algorithm~\cite{li_data-driven_2022}.

% 现有方法的不足
However, there remain some limitations in existing studies. First, the EM is a non-linear and non-additive system, indicating its parameters are coupled in a non-traceable way. Existing strategies to construct the identification set are either manual or numerical. However, a hybrid approach that takes advantage of both methods has not been tried yet. Second, in existing designs, the battery commonly needs to be tested under complicated or lengthy protocols. Some tests even rely on laboratory environments, which makes the identification time-consuming and impractical for implementation. Third, existing algorithms generally work in a stochastic manner, indicating a large number of times to repeatedly run the EM, which causes a considerable computation burden. In addition, different parameters are coupled together and impede the search for the optimum. To overcome these problems, an electrochemical parameter identification framework is proposed in this work. Parameters to be identified are determined and grouped manually first. Two sequenced tests are proposed to identify the specific parameters independently with a low time cost. Based on the result of numerical sensitivity analysis, a sensitivity-oriented stepwise optimization (SSO) algorithm is developed to search for the optimum efficiently. Case studies on a typical LIB type used in transportation energy systems are conducted for validation. There are three main contributions of this work.

\begin{itemize}
\item Electrochemical parameters are first grouped manually from the mechanism perspective and assigned to two sequenced tests for identification whose durations are compressed for practical implementation. Two optimization models are constructed respectively to simplify the solving procedure.
\item The Sobol method is applied to analyze the sensitivity of parameters and an SSO algorithm is developed to distinguish different parameters in the search for the optimum. By decomposing the solving process into preliminary and secondary stages, a proper solution can be found efficiently with low computation costs.
\item Numerical experiments on a typical LIB type widely used in transportation energy systems are conducted for validation. It is observed that the identified parameters can improve the accuracy of real-time battery monitoring and be used as indicators of the battery SOH. 
\end{itemize}

The rest of this paper is organized as follows: Section\ref{sec:test} introduces the identification model. Section\ref{sec:algor} presents the solving algorithm. Section\ref{sec:result} discusses the results of the numerical experiments. Section\ref{sec:con} draws conclusions.

\section{Identification Model}\label{sec:test}
In this section, electrochemical parameters to be identified are defined first. The test design and the optimization problem for identification are presented next.

\subsection{Parameter Definition}\label{sec:params}
\begin{figure*}[!htb]
	\centering
	\begin{minipage}{.95\textwidth}
		\centering
		\includegraphics[width=\textwidth]{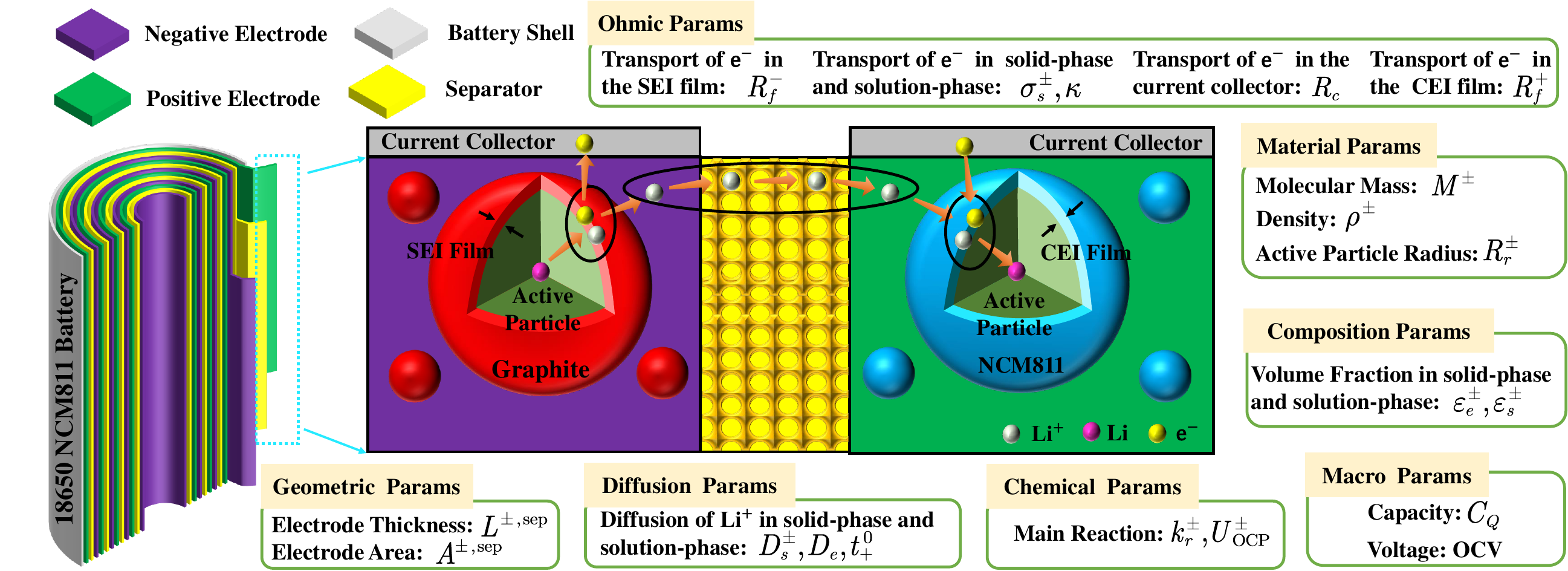}
	\end{minipage}
	\caption{Electrochemical parameters in a typical LIB.}\label{fig:diag}
\end{figure*}

To increase the practicability of this work, we start with an interpretable and analytical electrochemical model with low complexity~\cite{GU2023126192}. The parameters of the applied battery model are listed in Fig.~\ref{fig:diag}. 

As presented in Fig.~\ref{fig:diag}, electrochemical parameters can be grouped into seven categories.
\begin{itemize}
	\item The geometric parameters $A^{\pm,\mathrm{sep}}$ and $L^{\pm,\mathrm{sep}}$ are determined by the manufacturing. 
	\item The material parameters $M^{\pm}$, $\rho^{\pm}$ and $R_r^{\pm}$ reflect the physical property of active materials used in the electrodes. 
	\item The chemical parameters $k_r^{\pm}$ and $U_{\mathrm{OCP}}^{\pm}$ reflect the properties of the active material towards the Li-ion intercalation reaction. 
	\item Composition parameters $\varepsilon_{s}^{\pm}$ and $\varepsilon_{e}^{\pm}$ for a fresh battery are also determined by the manufacturing. However, some of them will decrease with the degradation of the battery. 
	\item Ohmic parameters $R_f^{\pm}$, $R_c$, $\sigma_s^{\pm}$ and $\kappa$ reflect the non-ideal transport of electrons in the solid electrolyte interphase (SEI) or catholyte electrolyte interphase (CEI), the current collector, the active particle and the electrolyte. It is noteworthy that SEI refers to the interphase in the negative electrode while CEI refers to that in the positive electrode~\cite{che_health_2023}.
	\item Diffusion parameters $D_s^{\pm}$, $D_e$ and $t_+^0$ depict the diffusion property of \ce{Li^+} in the active particle and electrolyte.  
	\item Macro parameters $C_Q$ and OCV curve reflect the output characteristics of the battery. 
\end{itemize}

For such a number of parameters, common identification approaches that optimize all parameters together are likely to obtain unreasonable results, as explained in the introduction. From the identification perspective, not every parameter needs to be identified. In this work, the parameters to be identified are determined manually based on their mechanisms.
\begin{itemize}
	\item The geometric parameters $A^{\pm,\mathrm{sep}}$ and $L^{\pm,\mathrm{sep}}$ are considered known because they can be obtained from the manufacturer and they will not change during the degradation.
	\item The material and chemical parameters $M^{\pm}$, $\rho^{\pm}$, $R_r^{\pm}$, $k_r^{\pm}$ and $U_{\mathrm{OCP}}^{\pm}$ reflect the inherent characteristics of active materials. For commonly used materials in commercial batteries, their values are available. In addition, they will not change during the degradation. Thus, they are considered known.
	\item Althouth the initial values of the composition parameters $\varepsilon_{s}^{\pm}$ and $\varepsilon_{e}^{\pm}$ can be obtained from the manufacturer. They might change during the degradation. Thus, they need to be identified. It is noteworthy that the changes in composition parameters are usually prolonged. 
	\item Among ohmic parameters, the SEI/CEI resistance $R_f^{\pm}$ are caused by the side reaction product accumulation. Their changing trajectories are directly related to the composition parameters. The rest parameters $R_c$, $\sigma_s^{\pm}$ and $\kappa$ are not explicitly related to the composition change, but can still vary with the SOH. Thus, they all need to be identified.
	\item Diffusion parameters are like ohmic parameters that can vary with the SOH and need to be identified.
	\item Macro parameters that reflect the external characteristics of the battery are derived from the above micro parameters after the identification.
\end{itemize}

\begin{figure}[!htb]
	\centering
	\begin{minipage}{.45\textwidth}
		\centering
		\includegraphics[width=\textwidth]{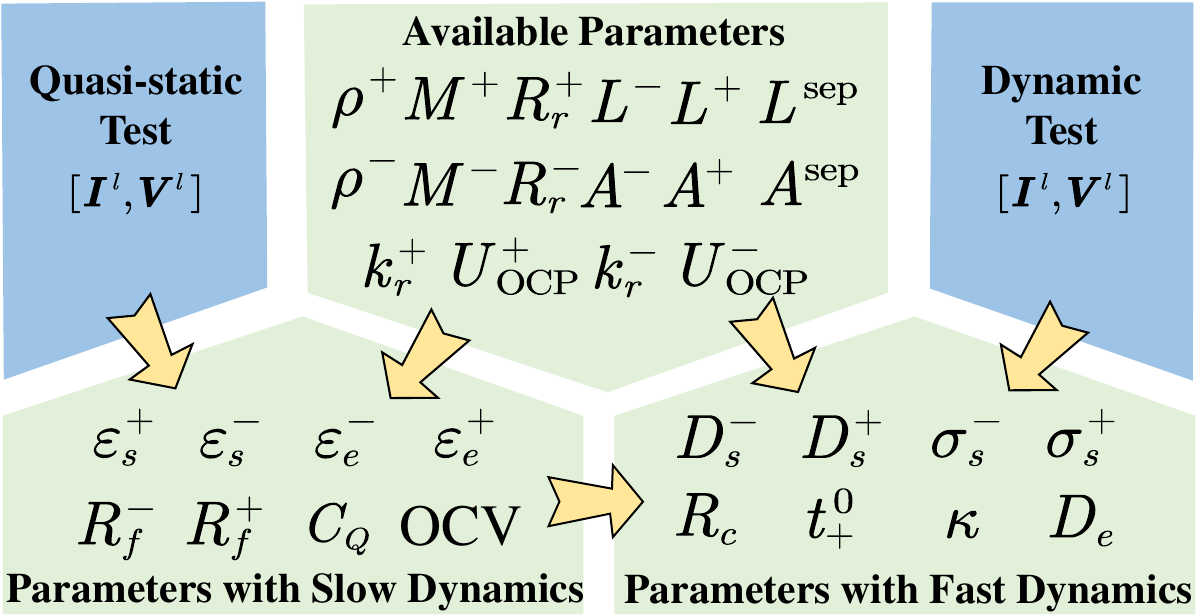}
	\end{minipage}
	\caption{The identification framework.}\label{fig:diag2}
\end{figure}

The entire identification procedure is decomposed into two steps to separate the contributions of different parameters. Specifically, two sequenced tests, the quasi-static and dynamic tests, are designed. In the quasi-static test, the battery is operated near a static status where only $\varepsilon_{s}^{\pm}$ are directly observable. In the dynamic test, the battery is operated at an active status where all parameters affect the voltage. The sketch of the entire identification framework is shown in Fig.~\ref{fig:diag2}. 

\subsection{Quasi-static Test}
\begin{figure}[!htb]
    \centering
    \begin{minipage}{.49\textwidth}
        \centering
        \includegraphics[width=\textwidth]{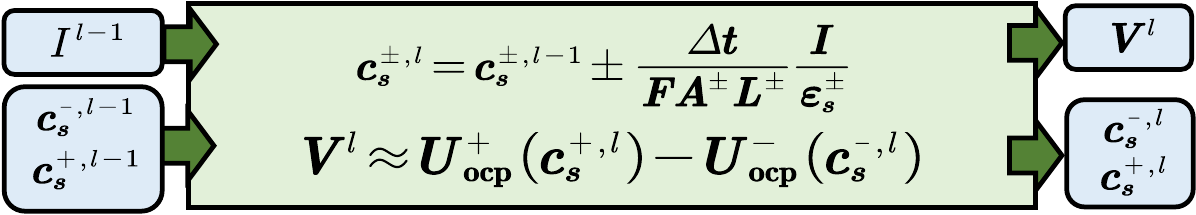}
    \end{minipage}
    \caption{The iterative step of a quasi-static battery model.}\label{fig:staticModel}
\end{figure}

As mentioned above, the battery works near a static status in the quasi-static test. Thus, a constant current with low amplitude is applied to the battery, suppressing other parameters' impacts. Under the quasi-static status, the ohmic over potential due to the non-ideal transport of electrons in different phases can be ignored. In addition, the diffusion of Li-ions in the solution-phase and solid-phase takes place slowly, inducing the Li-ion concentrations along the thickness direction in the electrolyte and the radius direction in the active particles to be approximately uniform. Thus, the polarization over potential due to the non-uniform distribution of Li-ions in different phases can be ignored. Under such circumstances, we do not need to know the values of ohmic and diffusion parameters in advance and the original battery model in ref.~\cite{GU2023126192} is simplified to the quasi-static form as given in Fig.~\ref{fig:staticModel}.
The first formula above is derived from the Coulomb counting method, i.e., the change of \ce{Li^+} concentrations in the solid-phase is proportional to the integral of the applied current in each step. The second formula is because over potentials caused by non-uniform distributions of concentrations are ignorable under low C-rate; thus the terminal voltage equals the solid-phase potential difference between the positive and negative electrodes. It is observed that only two composition parameters, $\varepsilon_s^{\pm}$, are involved. 

\begin{figure}[!htb]
    \centering
    \begin{minipage}{.49\textwidth}
        \centering
        \includegraphics[width=\textwidth]{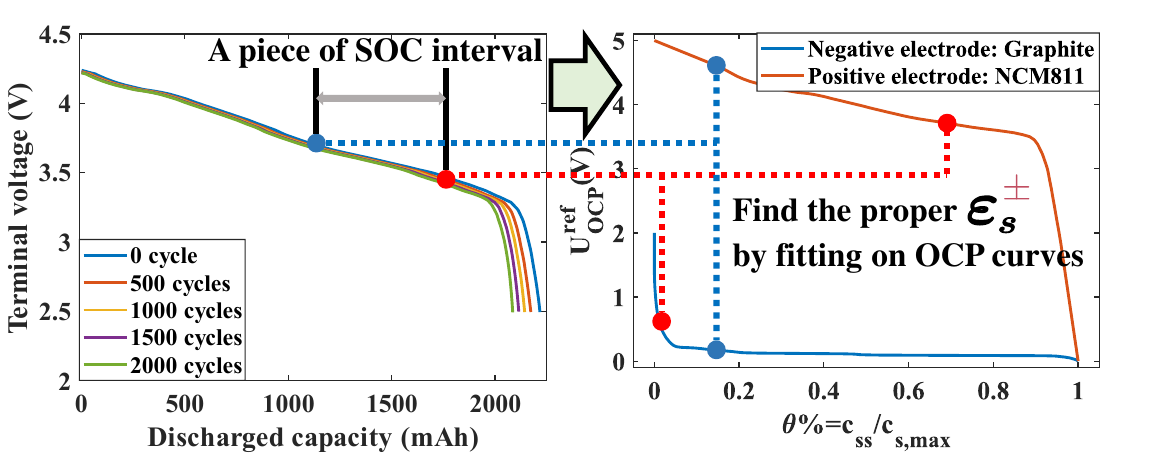}
    \end{minipage}
    \caption{The design and schematic diagram of the quasi-static test.}\label{fig:staticV}
\end{figure}

Generally, the quasi-static test is time-consuming due to the low current, e.g., it takes 100 hours to discharge the battery from full to empty when the current equals 0.01 C-rate. To save time, initial states of solid-phase \ce{Li^+} concentrations in active particles, denoted by $c_{s}^{\pm,0}$, are treated as pseudo parameters to identify. Thus, the test can start and stop at any point in the discharge curve as shown in Fig.~\ref{fig:staticV}.

Denote the measured voltage by $\boldsymbol{\hat{V}}$=$[\hat{V}^1,\cdots,\hat{V}^N]$ and the static battery model by $f_{\mathrm{static}}$, the optimization problem for identification is formulated as:
\begin{equation}
    \label{equ:staticOpt}
    \begin{split}
    &\min_{[c_{s}^{\pm,0}, \varepsilon_s^{\pm}]} \quad  \sum_{l=1}^N{\left( \hat{V}^l -V^l \right) ^2}, 
    \\
    &\mathrm{s}.\mathrm{t}. \left[ V^l , c_{s}^{\pm,l} \right] =f_{\mathrm{static}}\left(  c_{s}^{\pm,l-1}; I,\varDelta t ,\varepsilon _{s}^{\pm} \right).
    \end{split}
\end{equation}
The optimization problem above aims to minimize the voltage error between the measurements and the model output. The constraint is the model computation process.

After obtaining $\varepsilon_s^{\pm}$ and $c_{s}^{\pm,0}$, macro-scope characteristics of the battery, i.e., the battery capacity $C_Q$ and the open circuit voltage (OCV), can be directly derived. In real-world applications, the operating region of the battery is usually depicted by the voltage limitation suggested by the manufacturer. Denote the maximum and minimum approved voltages by $V_{\mathrm{max}}$ and $V_{\mathrm{min}}$, the limitation of $c_s^{\pm}$ when the battery is discharged to empty, denoted by $c_{s,\mathrm{max}}^+$ and $c_{s,\mathrm{min}}^-$, can be solved by:
\begin{equation}
    \label{equ:thetalimit}
    \begin{cases}
    \left( c_{s,\mathrm{max}}^+ - c_s^{+,0} \right)\gamma=\left(c_s^{-,0}-c_{s,\mathrm{min}}^-\right)
    \\
    U_{\mathrm{OCP}}^+(c_{s,\mathrm{max}}^+) - U_{\mathrm{OCP}}^-(c_{s,\mathrm{min}}^-) = V_{\mathrm{min}}
    \end{cases} \Rightarrow 
	\begin{cases}
		c_{s,\mathrm{max}}^+ \\ c_{s,\mathrm{min}}^-
	\end{cases}
\end{equation}
where $\gamma=\frac{\varepsilon_s^+A^+L^+}{\varepsilon_s^-A^-L^-}$ is the volume ratio of positive electrode active materials to negative electrode active materials. The first equation is derived from the ionic conservation law. When the battery is discharged to empty at the current state, cyclable Li-ions move from the negative electrode to the positive electrode, and the quantity of Li-ions deintercalated from the negative electrode equals the quantity of  Li-ions intercalated into the positive electrode. The second equation is derived from the voltage limitation specified by the user, i.e. when a battery is defined as empty, it means its voltage reaches $V_{\mathrm{min}}$ rather than indicating there are no Li-ions in the negative electrode.

Similarly, the limitation of $c_s^{\pm}$ when the battery is charged to full, denoted by $c_{s,\mathrm{min}}^+$ and $c_{s,\mathrm{max}}^-$, can be solved by replacing $c_{s,\mathrm{max}}^+$, $c_{s,\mathrm{min}}^-$, and $V_{\mathrm{min}}$ in Eq. (\ref{equ:thetalimit}) with $c_{s,\mathrm{min}}^+$, $c_{s,\mathrm{max}}^-$, and $V_{\mathrm{max}}$, respectively. Based on $c_{s,\mathrm{min}}^{\pm}$, $c_{s,\mathrm{max}}^{\pm}$, the total capacity of the battery equals the sum of the maximum capacity that can be injected and extracted at the current state as follows:
\begin{equation}
    \label{equ:cellCapacity}
    C_Q = \left(c_{s,\mathrm{max}}^- - c_{s,\mathrm{min}}^- \right)\varepsilon_s^-A^-L^-.  
\end{equation}
For any $\mathrm{SOC}_i\in[0,1]$, the corresponding OCV is as follows:
\begin{equation}
    \label{equ:SOC-OCV}
    \begin{split}
       \mathrm{OCV}\left(\mathrm{SOC}_i\right) &= U_{\mathrm{OCP}}^+\left(c_{s,\mathrm{max}}^+ - \mathrm{SOC}_i  \left(  c_{s,\mathrm{max}}^+ - c_{s,\mathrm{min}}^+ \right) \right) \\ & - U_{\mathrm{OCP}}^-\left(c_{s,\mathrm{min}}^- + \mathrm{SOC}_i  \left(  c_{s,\mathrm{max}}^- - c_{s,\mathrm{min}}^- \right) \right). 
    \end{split}
\end{equation}

Although the electrolyte volume fraction $\varepsilon_e^{\pm}$ and SEI/CEI resistance $R_f^{\pm}$ are not directly reflected in the voltage, an empirical model based on the degradation mechanism inside the battery is constructed to identify them in this stage instead of leaving them to the next test. 

Take the commonly-used NCM811 battery as an example (the positive electrode is made of \ce{Li(Ni_{0.8}Co_{0.1}Mn_{0.1})O_2} active particles), the chemical formulas of the SEI and CEI formation are expressed as: 2\ce{Li}+\ce{2C_2H_4CO_3}\ce{=}\ce{(CH_2 O C O_2Li)_2}+\ce{C_2H_4};  \ce{Li(Ni_{0.8}Co_{0.1}Mn_{0.1})O_2}+\ce{2C_2H_4CO_3}\ce{=}\ce{(Ni_{0.8}Co_{0.1}Mn_{0.1})O_2CH_2}+\ce{CH_2OCO_2Li}. 
With the growth of SEI/CEI, the film over the active particle becomes thicker, leading to the increment of $R_f^{\pm}$. Since the composition of SEI/CEI is known, the electrical conductivity, denoted by $\sigma_f^{\pm}$, is assumed to be known. According to Ohm's law, $R_f^{\pm}$ can be expressed by $R_f^{\pm}$=$\Delta \delta_f^{\pm}/\sigma_f^{\pm}$, i.e., knowing the SEI/CEI thickness $\delta_f^{\pm}$ is the key to identifying $R_f^{\pm}$. 

The chemical formula indicates that the CEI formation in the positive electrode consumes the active particles in proportion. Thus, the relationship between the increment of $\delta_f^+$ and the decrement of $\varepsilon_s^+$ is fitted by the linear mapping. However, it is not the case for the negative electrode since the SEI formation does not involve the graphite active particles. According to ref.~\cite{zhao_electrochemical-thermal_2019}, the thickening of SEI will occupy the space of electrolyte due to its microstructure and lead to the decrement of $\varepsilon_e^-$. The above research also suggested that the relationship between the increment of $\delta_f^-$ and the decrement of $\varepsilon_e^-$ follows the linear mapping. The conclusion is adopted in this work. Thus, we have:
\begin{equation}
    \label{equ:electrode_aged}
     \Delta \delta_f^+ = k_f^+ \Delta \varepsilon_{s}^+, \quad \Delta \delta_f^- =  k_f^- \Delta  \varepsilon_{e}^-.
\end{equation}
where $k_f^{\pm}$ are fitted coefficients. In the formula above, $\varepsilon_e^-$ is still unknown. Since $\varepsilon_s^-$ can be identified, an empirical model is built to capture the relationship between $\Delta \varepsilon_e^-$ and $\Delta \varepsilon_s^-$. Since the previously formed SEI can prevent subsequent SEI growth, the SEI formation's reaction rate will decrease with the progress of aging~\cite{che_health_2023}. Considering this rule, the relationship between the decrement of $\varepsilon_e^-$ and the decrement of $\varepsilon_s^-$ is fitted by the quadratic mapping:
\begin{equation}
	\label{equ:C6epsilone}
	\Delta \varepsilon_e^- = k_e^- \left( \Delta \varepsilon_s^- \right)^2 + b_e^- \Delta \varepsilon_s^-. 
\end{equation}
where $k_e^-$ and $b_e^-$ are fitted coefficients. It is noteworthy that the linear approximation in Eq.(\ref{equ:electrode_aged}) and quadratic approximation in Eq.(\ref{equ:C6epsilone}) are both verified in the experimental data obtained from the AutoLion~\cite{AutoLion}. To note, the change of $\varepsilon_e^+$ is neglected because there is no report in existing studies that CEI will cause the decrement of porosity (i.e., the $\varepsilon_e^+$), the experimental data from AutoLion also verify this phenomenon.

To conclude, $\varepsilon_{s}^{\pm}$, $R_f^{\pm}$, $\varepsilon_{e}^-$, $C_Q$ and OCV can be identified in the quasi-static test. Typically, these parameters are all directly related to the fade of material composition.

\subsection{Dynamic Test}\label{sec:dynamic test}
\begin{figure}[!htb]
	\centering
	\begin{minipage}{.49\textwidth}
			\centering
			\includegraphics[width=\textwidth]{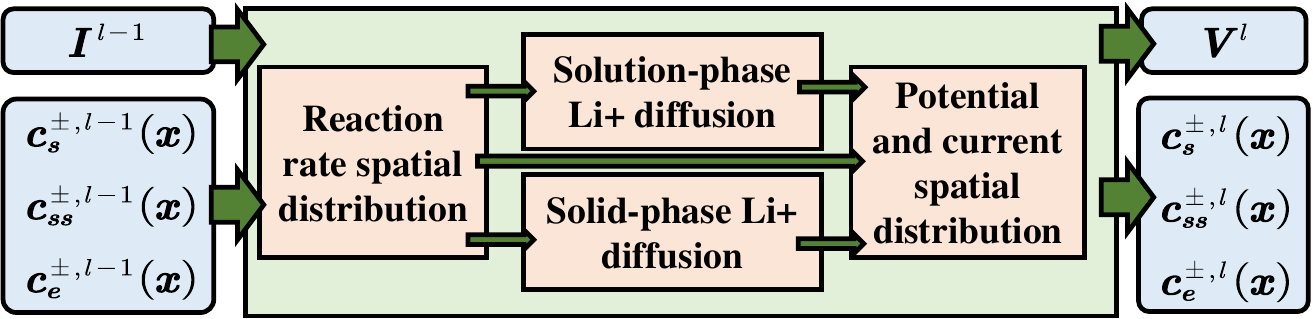}
	\end{minipage}
\caption{The iterative step of the battery model in ref.~\cite{GU2023126192}.}\label{fig:dynamicModel}
\end{figure}

In the dynamic test, the battery is operated in an active status so that impacts of all parameters can be reflected in the voltage. Under such circumstances, the nonuniform distributions of reaction rates, \ce{Li^+} concentrations, potentials, and currents should be considered. Thus, the complete battery model in ref.~\cite{GU2023126192} is adopted. As shown in Fig.~\ref{fig:dynamicModel}, unlike the quasi-static model, the concentration states are expressed as functions of the spatial coordinates $x$.

The rest parameters to be identified in Fig.~\ref{fig:diag2} can be further categorized into two groups. The first group contains $\kappa$, $\sigma_s^{\pm}$ and $R_c$. Physically, they are related to the transport of electrons, indicating their impacts are instantly reflected in the voltage when the current suddenly changes. The second group contains $D_s^{\pm}$, $D_e$ and $t_+^0$. They are related to the transport of Li-ions, indicating their impacts are gradually reflected in the voltage until the battery reaches a steady point. This phenomenon is because the transfer speed of electrons is much faster than ions. 

\begin{figure}[!htb]
	\centering
	\begin{minipage}{.49\textwidth}
		\centering
		\includegraphics[width=\textwidth]{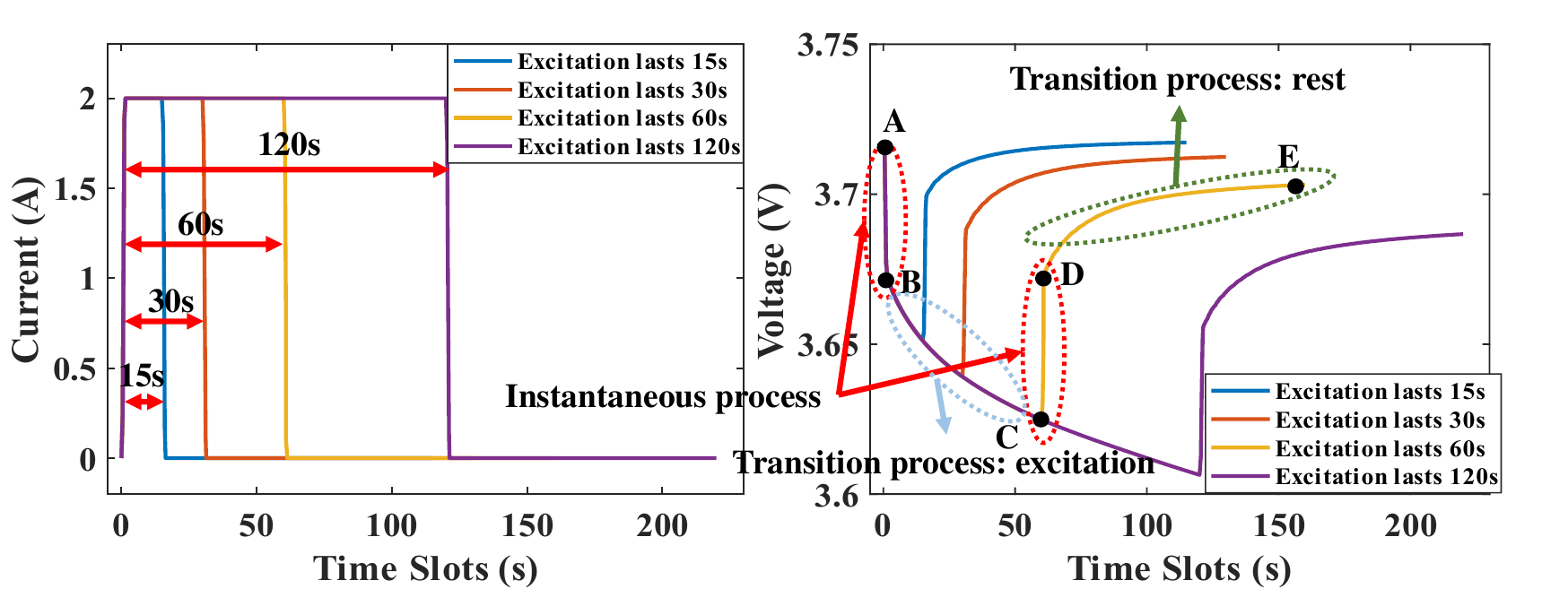}
	\end{minipage}
	\caption{The design and schematic diagram of the dynamic test.}\label{fig:dynamicV}
\end{figure}

Considering the above rule, four sequenced square-wave pulses with durations ranging from 115 seconds to 220 seconds are adopted as the input current during the test, as shown in the left of Fig.~\ref{fig:dynamicV}. For each square-wave pulse, i), the duration of the excitation process, ii), the duration of the rest process, iii), the amplitude of the excitation current, iv), the battery SOC at the start of the test, should be determined in advance. 

The duration of the excitation process and the rest process are determined first. According to the model in ref.~\cite{GU2023126192}, the time constants of first-order processes controlled by specific parameters are different. To separate the impacts of different parameters, the durations of the excitation process for four pulses are set to 15, 30, 60, and 120 seconds, respectively. By cutting off the current after different lasting times, the proportions of different first-order processes in the voltage can be manipulated. For example, the long-time-constant processes will be partially recorded under short excitation and fully recorded under prolonged excitation. In this way, the parameters that are related to different first-order processes can be distinguished better. According to numerical experiments, it takes about 120 seconds (as shown in the right of Fig.~\ref{fig:dynamicV}) for the battery to reach a steady status under the galvanostatic current, i.e., the voltage gradient on time gradually approaches a constant. Thus, the maximum duration of the excitation is set to 120 seconds. The minimum duration of the excitation is set to 15 seconds to ensure the measurement data is sufficient for analysis. In the rest period, the duration time should give the battery enough time to reach a steady status, i.e., the voltage becomes stable. According to numerical experiments, it takes around 100 seconds (as shown in the right of Fig.~\ref{fig:dynamicV}) for the voltage to converge to be constant. Thus, the duration of the rest is set to 100 seconds. 

Next, the current amplitude and the SOC starting point should be determined. Through numerical experiments, we find that these two parameters have little influence on the identification accuracy. Thus, without loss of generality, the current amplitude is set to 1 C-rate, and the initial SOC is set to $0.6$. Summing the durations of the four sequenced square-wave pulses, the whole dynamic test takes $625$ seconds in all. 

The voltage response of each square-wave pulse is shown in the right of Fig.~\ref{fig:dynamicV}, it typically contains four parts separated by points A,$\cdots$,E: i), zero-to-step instantaneous process at the start of the excitation (i.e., between point A and B), ii), zero-to-step transient process until the battery reaches an active-steady status under the galvanostatic current (i.e., between point B and C), iii), step-to-zero instantaneous process at the end of the excitation (i.e., between point C and D), iv), step-to-zero transient process until the battery reaches a static-steady status under the rest (i.e., between point D and E). 

The basic idea of the identification in this stage is to assign parameters to specific voltage segments for optimization. Typically, parameters that immediately take effect will be identified by the optimization over the instantaneous process. Parameters that dominate first-order processes inside the battery will be identified by the optimization over the transient process, either the excitation period or the rest period. 

Denote the time indexes of points A,$\cdots$,E in the voltage series by $l_{\mathrm{A}}$,$\cdots$,$l_{\mathrm{E}}$. For the set of parameters optimized over the instantaneous process, denoted by $\boldsymbol{\Theta}^{\mathrm{I}}$, the corresponding objective function, denoted by $f_{\mathrm{I}}$, is expressed by:
\begin{equation}
	\label{equ:obj_i}
	f_{\mathrm{I}} = \sum_{l=l_{\mathrm{A}}}^{l_{\mathrm{B}}} {\left( \hat{V}^l - V^l \right) ^2} + \sum_{l=l_{\mathrm{C}}}^{l_{\mathrm{D}}} {\left( \hat{V}^l -V^l \right) ^2}.
\end{equation}
Obviously, $f_{\mathrm{I}}$ is equivalent to the root mean square error (RMSE) of the battery voltage in the instantaneous process.

Similarly, for sets of parameters optimized over the excitation process and rest process, denoted by $\boldsymbol{\Theta}^{\mathrm{E}}$ and $\boldsymbol{\Theta}^{\mathrm{R}}$, the corresponding objective functions $f_{\mathrm{E}}$ and $f_{\mathrm{R}}$ are expressed by:
\begin{equation}
	\label{equ:obj_e}
	f_{\mathrm{E}} = \sum_{l=l_{\mathrm{B}}}^{l_{\mathrm{C}}}\left(\left(\hat{V}^l - \hat{V}^{l_{\mathrm{B}}}\right)-\left(V^l-V^{l_{\mathrm{B}}}\right)\right)^2.
\end{equation}
\begin{equation}
	\label{equ:obj_r}
	f_{\mathrm{R}} = \sum_{l=l_{\mathrm{D}}}^{l_{\mathrm{E}}}\left(\left(\hat{V}^l - \hat{V}^{l_{\mathrm{D}}}\right)-\left(V^l-V^{l_{\mathrm{D}}}\right)\right)^2.
\end{equation}
It is noteworthy that the bias originating from the instantaneous process is eliminated in the above two formulas to prevent the impact of $f_{\mathrm{I}}$.

In Eqs.(\ref{equ:obj_i})-(\ref{equ:obj_r}), the voltage series $V^l$ is obtained by the battery model proposed in ref.~\cite{GU2023126192}, denoted by $f_{\mathrm{dynamic}}$. Denote the internal states at the $l$-th step by $\boldsymbol{c}^l$=$\left[ c_{s}^{\pm,l}\left(x\right), c_{ss}^{\pm,l}\left(x\right), c_{e}^{\pm,l}\left(x\right) \right]$. The optimization problem can be expressed by:
\begin{equation}
	\label{equ:dynamicOpt}
	\begin{split}
		&\min_{\boldsymbol{\Theta}^{\mathrm{I/E/R}}} \quad  f_{\mathrm{I/E/R}}, 
		\\
		&\mathrm{s}.\mathrm{t}. \left[V^l, \boldsymbol{c}^l\right] =
		f_{\mathrm{dynamic}}\left( \boldsymbol{c}^{l-1}, I^l; \Delta t, \boldsymbol{\Theta}^{\mathrm{I}},\boldsymbol{\Theta}^{\mathrm{E}},\boldsymbol{\Theta}^{\mathrm{R}} \right).
	\end{split}
\end{equation}
In the formula above, the objective and the optimization variable are optional. For example, when identify parameters in $\boldsymbol{\Theta}^{\mathrm{I}}$, $f_{\mathrm{I}}$ is used as the objective function. However, such a design makes the problem intractable, e.g., when identifying parameters in $\boldsymbol{\Theta}^{\mathrm{I}}$, parameters in $\boldsymbol{\Theta}^{\mathrm{E}}$ and $\boldsymbol{\Theta}^{\mathrm{R}}$ need to be known to obtain the constraint in Eq.(\ref{equ:dynamicOpt}), which are unknown. Thus, a specific algorithm needs to be developed to solve the problem.

It is also noteworthy that $D_s^{\pm}$, $D_e$ and $\kappa$ will vary with the temperature or reactant concentrations~\cite{torchio_lionsimba_2016}. To improve the accuracy, these four parameters are expressed in the function form~\cite{GU2023126192}. Directly identifying their expressions is intractable. Thus, correction factors are introduced to transform them into the scalar format. For example, $\kappa$=$\hat{\kappa}$$f_{\kappa}$, where $f_{\kappa}$ is a functional expression and $\hat{\kappa}$ is the correction factor. By replacing $\kappa$ with $\hat{\kappa}$ in the identification task, the basic properties of $\kappa$ are retained while the degradation trajectory of $\kappa$ can be identified as well. For notation simplicity, ($\cdot$) and ($\hat{\cdot}$) are not distinguished in the rest of this article.

\section{Algorithm}\label{sec:algor}
In this section, applied algorithms for identifying parameters in the quasi-static and dynamic tests are designed.

\subsection{Classic heuristic algorithms}
In the quasi-static test, $\varepsilon_s^{\pm}$ and $c_s^{\pm,0}$ are identified by solving Eq.(\ref{equ:staticOpt}). As shown in the right of Fig.~\ref{fig:staticModel}, $U_{\mathrm{OCP}}^{\pm}$ are nonlinear, indicating that exact optimization algorithms are likely to be trapped in a local minimum when encountering incorrect initialization. Thus, classic heuristic optimization algorithms are applied, which are efficient in searching for the optimum of a non-linear problem, i.e., PSO or GA. The basic idea of the PSO and GA is searching for the optimum via the competition and cooperation of populations under a stochastic framework. It is noteworthy that heuristic algorithms commonly consume more computation sources than exact algorithms. However, in the quasi-static identification task, the time cost of heuristic algorithms is acceptable for two reasons. First, the complexity of a quasi-static battery model is low as shown in Fig.~\ref{fig:staticModel}. Second, the quasi-static test is only carried out in a partial SOC range. In addition, the voltage changes slowly under quasi-static conditions, so the sampling frequency is low. Thus, there are few data points under quasi-static conditions. Combining these two points, heuristic algorithms can give solutions efficiently as well.

Classic heuristic algorithms can be conveniently called on different computation platforms, e.g., MATLAB and Python. For details of these two algorithms, refer to refs.~\cite{785511,katoch2021review}.

\subsection{Sensitivity-oriented stepwise optimization (SSO) algorithm}
Denote the set of parameters to be identified in the dynamic test by $\boldsymbol{\Theta}$, their values are obtained by solving Eq.(\ref{equ:dynamicOpt}). As analyzed above, existing solvers cannot tackle the problem directly. Thus, an SSO algorithm composed of a three-stage solving procedure is designed: i), analyzing the sensitivity of each parameter to voltage segments obtained in the dynamic test and separating $\boldsymbol{\Theta}$ into three sets: $\boldsymbol{\Theta}^{\mathrm{I}}$, $\boldsymbol{\Theta}^{\mathrm{R}}$ and $\boldsymbol{\Theta}^{\mathrm{E}}$; ii), identifying parameters in $\boldsymbol{\Theta}^{\mathrm{I}}$ in a stepwise manner; iii), identifying parameters in $\boldsymbol{\Theta}^{\mathrm{T}}$=$\boldsymbol{\Theta}^{\mathrm{E}} \cup \boldsymbol{\Theta}^{\mathrm{R}}$ in a stepwise manner.

\subsubsection{Sensitivity analysis}
Under the active status, the battery can be viewed as a non-additive and nonlinear system, i.e., the voltage is attributed not only to each parameter independently but also to their interactions. Thus, a global sensitivity analysis framework, i.e., the Sobol method, is applied~\cite{sobol1990sensitivity}. The Sobol method works within a probabilistic framework and is suitable for multi-parameter systems.

Take the sensitivity of the $k$-th parameter in $\boldsymbol{\Theta}$ to a specific voltage segment as an example. First, introduce the concept of parameter random-variables: $\boldsymbol{\theta}_{k}$ means only the $k$-th element in $\boldsymbol{\Theta}$ is a random-variable and $\boldsymbol{\theta}_{\sim k}$ means all elements except the $k$-th element in $\boldsymbol{\Theta}$ is a random-variable. Substituting a parameter random-variable into Eq.(\ref{equ:dynamicOpt}) and calculating the corresponding objective function yields a scalar output, denoted by $F$, which is also a random variable. The Sobol sensitivity can be calculated by:
\begin{equation}
	\label{equ:sobol_formula}
	s_k = \frac{{\rm E}_{\boldsymbol{\theta}_{\sim k}}\left( {\rm Var}_{\boldsymbol{\theta}_k} \left( F|\boldsymbol{\theta}_{\sim k} \right) \right)}{{\rm{Var}}(F)}. 
\end{equation}

Mathematically, $s_k$ represents the fraction of the total output variance that would remain on average as long as the $k$-th parameter stays unknown. A higher $s_k$ indicates that the $k$-th parameter is more sensitive to the voltage segment. For a complex system, calculating Eq.~(\ref{equ:sobol_formula}) by analytical integrals is intractable. The Sobol method gives a numerical way to estimate $s_{k}$ based on Monte Carlo integrals. As the sampling scale increases, the Monte Carlo results gradually converge to the analytical results. According to existing research, replacing conventional random sequences with quasi-random low-discrepancy sequences such as the Sobol sequence can further improve the convergence rate~\cite{zadeh_comparison_2017}, which is adopted in this work. For details of generating the Sobol random numbers, refer to ref.~\cite{burhenne2011sampling}.

To compute the Sobol sensitivity numerically, the sampling interval of each parameter should be determined first. Generally, a wider interval can give more reliable results. By applying the aforementioned  quasi-Monte Carlo sequence on the defined intervals, $2M$ samples of $\boldsymbol{\Theta}$ are generated, compact them to the matrix form:
\begin{equation}
	\boldsymbol{\Xi}=\left[ \begin{array}{c}
		\boldsymbol{\Theta }_{1}\\
		\boldsymbol{\Theta }_{2}\\
		\vdots\\
		\boldsymbol{\Theta }_{2M}\\
	\end{array} \right] =\left[ \begin{matrix}
		\theta _{1,1}&		\theta _{1,2}&		...&		\theta _{1,p}\\
		\theta _{2,1}&		\theta _{2,2}&		...&		\theta _{2,p}\\
		\vdots&		\vdots&		\ddots&		\vdots\\
		\theta _{2M,1}&		\theta _{2M,2}&		\cdots&		\theta _{2M,p}\\
	\end{matrix} \right] _{2M\times p}
\end{equation}
where $p$ is the number of parameters in $\boldsymbol{\Theta}$. Every row of $\boldsymbol{\Xi}$ corresponds to an independent sample. Substituting the $j$-th row into Eq.(\ref{equ:dynamicOpt}) and calculating the objective function, denoted by $f_{j}$. The denominator of Eq.(\ref{equ:sobol_formula}) can be estimated by:
\begin{equation}
	\label{equ:var_Y}
	{\rm Var}(F) \approx \frac{1}{2M-1}\sum_{j=1}^{2M}(f_{j}-\bar{f})^2.
\end{equation}
where $\bar{f}$ equals the mean value of $f_{j}$ for $j$=$1,2,\cdots,2M$.

To calculate the numerator of (\ref{equ:sobol_formula}), we construct a new $M$$\times$$p$ matrix $\boldsymbol{\Xi}^k$ based on $\boldsymbol{\Xi}$:
\begin{equation}
    \boldsymbol{\Xi}^k=\left[ \begin{array}{c}
	{\boldsymbol{\Theta }_{k,1}}\\
	{\boldsymbol{\Theta }_{k,2}}\\
	\vdots\\
	{\boldsymbol{\Theta }_{k,M}}\\
\end{array} \right] =\left[ \begin{matrix}
	\theta _{1,1}&		\begin{matrix}
	\cdots&		\theta _{M+1,k}\\
\end{matrix}&		\cdots&		\theta _{1,p}\\
	\theta _{2,1}&		\begin{matrix}
	\cdots&		\theta _{M+2,k}\\
\end{matrix}&		\cdots&		\theta _{2,p}\\
	\vdots&		\begin{matrix}
	\ddots&		\vdots\\
\end{matrix}&		\ddots&		\vdots\\
	\theta _{M,1}&		\begin{matrix}
	\cdots&		\theta _{2M,k}\\
\end{matrix}&		\cdots&		\theta _{M,p}\\
\end{matrix} \right] _{M\times p}
\end{equation}
Specifically, the $k$-th elements of the 1-st to $M$-th rows of $\boldsymbol{\Xi}$ are replaced by the $k$-th elements of the $(M$$+1)$-th to $2M$-th rows of $\boldsymbol{\Xi}$. Substituting the $j$-th row into Eq.(\ref{equ:dynamicOpt}) and calculating the objective function denoted by $f_{k,j}$. The numerator of Eq.(\ref{equ:sobol_formula}) can be estimated by:
\begin{equation}
    {\rm E}_{\boldsymbol{\theta}_{\sim k}}\left( {\rm Var}_{\boldsymbol{\theta}_k} \left( F|\boldsymbol{\theta}_{\sim k} \right) \right) \approx \frac{1}{2M}\sum_{j=1}^M\left(  f_{j}-f_{k,j} \right)^2.
\end{equation}

Denote the voltage segments of the instantaneous process in pulses with durations from 15 to 120 seconds by $\zeta_1$$\sim$$\zeta_4$, the voltage segments of the excitation process by $\zeta_5$$\sim$$\zeta_8$, the voltage segments of the rest process by $\zeta_9$$\sim$$\zeta_{12}$.

According to Fig.~\ref{fig:diag2}, there are $8$ parameters to be identified in this stage. The Sobol sensitivity of every parameter to every voltage segment is computed and arranged to an 8$\times$12 matrix $\boldsymbol{S}$ as shown in Fig.~\ref{fig:Algorithm}, the element on the $k$-th row and $j$-th column, $s_k^j$ refers to the sensitivity of the $k$-th parameter to the $j$-th voltage segment. For the $k$-th parameter, calculate the mean sensitivity to voltage segments of the instantaneous process, excitation process and rest process, denoted by $\bar{s}^{\mathrm{I}}_k$, $\bar{s}^{\mathrm{E}}_k$ and $\bar{s}^{\mathrm{R}}_k$, respectively. It is assigned to $\boldsymbol{\Theta}^{\mathrm{I}}$ if $\bar{s}^{\mathrm{I}}_k$ is the maximum, and assigned to $\boldsymbol{\Theta}^{\mathrm{T}}$ if not.

\subsubsection{Stepwise optimization}
Considering the measurement precision of voltage sensors, parameters with the Sobol sensitivity below $0.01$ are removed from the identification task and directly set to empirical values. Parameters in $\boldsymbol{\Theta}^{\mathrm{I}}$ are identified first because the effects of parameters in $\boldsymbol{\Theta}^{\mathrm{T}}$ have not yet been reflected in the voltage. Denote the size of $\boldsymbol{\Theta}^{\mathrm{I}}$ by $N_\mathrm{I}$, resorting its elements by $\bar{s}^{\mathrm{I}}_k$ in the descending order yield $\boldsymbol{\Theta}^{\mathrm{I}}$=$\{\theta_1, \theta_2, \cdots, \theta_{N_\mathrm{I}}\}$. 

First, the preliminary estimation is conducted one by one from $\theta_1$ to $\theta_{N_{\mathrm{I}}}$. Taking $\theta_1$ as an example, locate the voltage segment $\zeta_j$ where it has the highest sensitivity, i.e., $j$=$\arg\mathrm{max}$ $s^j_1$, $j$=$1,2,3,4.$ The preliminary estimation, denoted by $\check{\theta}_1$, is obtained by minimizing $f_{\mathrm{I}}$ in $\zeta_j$. Since $\theta_2$$\sim$$\theta_8$ are unknown, they are set to random values between reasonable ranges. The adopted optimization algorithm can be either exact algorithms, i.e., the active set method, or the heuristic algorithms, i.e., PSO or GA. The selection of the algorithm is determined by the effect of $\theta_1$ on the voltage. Specifically, suppose the mapping between $\theta_1$ and $f_{\mathrm{I}}$ is approximately convex. In that case, the exact algorithm with high speed is selected since the risk of being trapped in a local minimum is low. Otherwise, the heuristic algorithm is selected. 
Due to the complexity of the battery model, explicitly analyzing the convexity of the mapping between a given parameter and the voltage is intractable. In this work, the selection between the exact and heuristic algorithms is determined manually based on the results of numerical experiments. For the subsequent parameters, repeat the steps above. The only difference is that when estimating the $i$-th parameter $\theta_i$, set $\theta_j$=$\check{\theta}_j$, $j$$<$$i$ and $\theta_j$, $j$$>$$i$ to random values. Finally, the preliminary estimations of all parameters in $\boldsymbol{\Theta}^{\mathrm{I}}$ are known, denoted by $\{ \check{\theta}_1, \check{\theta}_2, \cdots, \check{\theta}_{N_{\mathrm{I}}} \}$. 

Next, the voltage piece $\zeta_j$ is located such that the difference between the highest sensitivity and the lowest sensitivity is smallest, i.e., $j$=$\arg\mathrm{min} \left(  \mathrm{max} (s^j_m) - \mathrm{min} (s^j_n)   \right)$, $j$=$1,2,3,4.$, and $m,n$=$1,2,$$\cdots$,$N_{\mathrm{I}}$. This ensures the weights of all parameters are relatively equal. Then, the final identification is obtained by optimizing $f_{\mathrm{I}}$ over the voltage segment $\zeta_j$. Since all parameters are involved in this step, the heuristic algorithm is adopted to search for the global optimum. The lower and the upper bounds of optimization variables are set as 95\% and 105\% of the preliminary estimation values, i.e., $0.95\check{\theta}_j$ $\le$ $\theta_j$ $\le$ $ 1.05\check{\theta}_j$, $j$=$1,2,$$\cdots$,$N_{\mathrm{I}}$, which ensures that the identification of parameters with low sensitivity is not being affected by other parameters considerably. The result obtained in this step is treated as the final estimation, denoted by $\{ \hat{\theta}_1, \hat{\theta}_2, \cdots, \hat{\theta}_{N_{\mathrm{I}}} \}$.

\begin{figure*}[!htb]
	\centering
	\begin{minipage}{.8\textwidth}
		\centering
		\includegraphics[width=\textwidth]{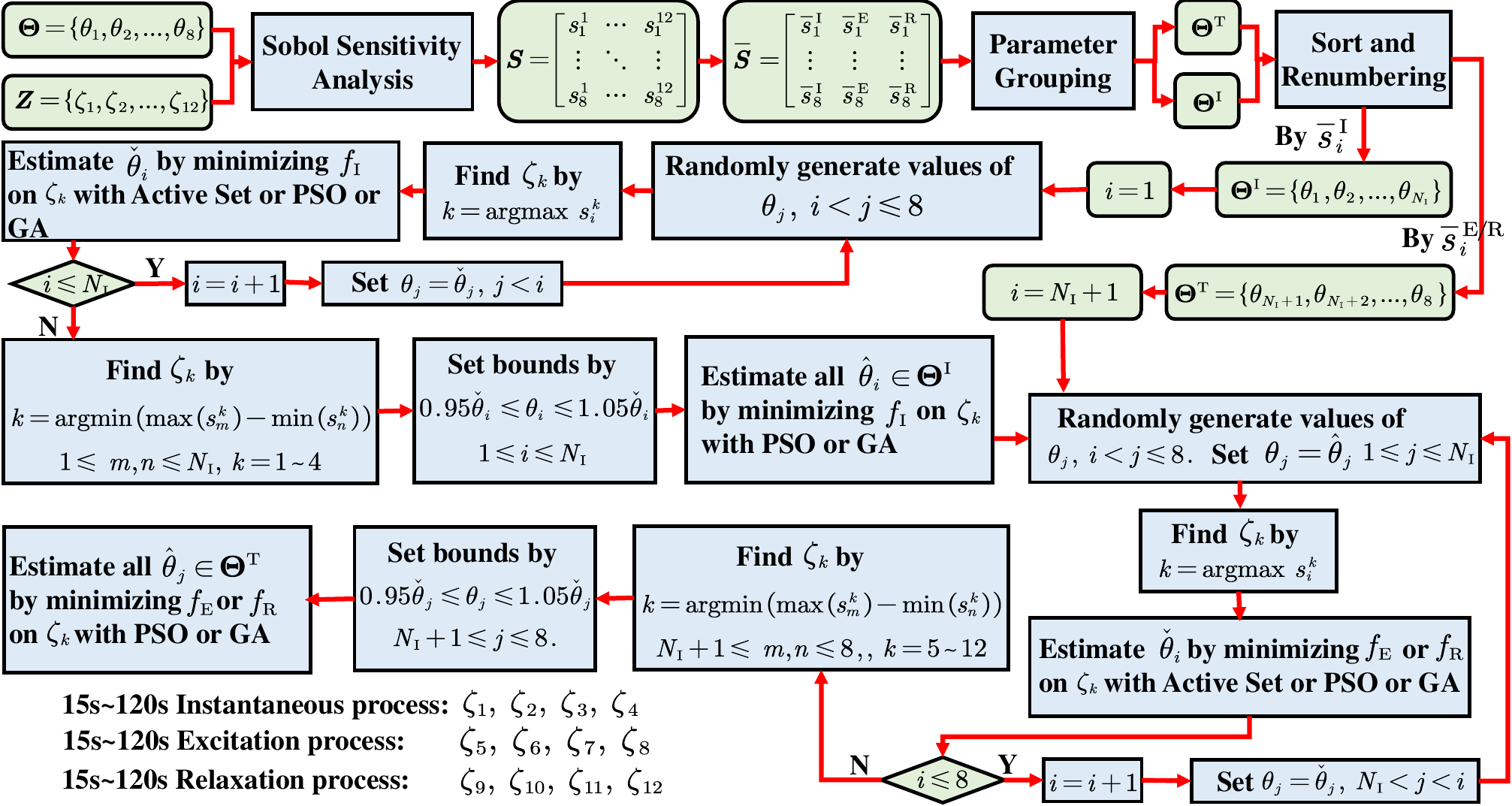}
	\end{minipage}
	\caption{The SSO algorithm prosedure. }\label{fig:Algorithm}
\end{figure*}

Parameters in $\boldsymbol{\Theta}^{\mathrm{T}}$=$\{ \theta_{N_{\mathrm{I}}+1}, \theta_{N_{\mathrm{I}}+2},\cdots,\theta_8
\}$ are identified similarly. First, sort the parameters according to their sensitivity in the descending order. Second, conduct the preliminary estimation for each parameter over the proper voltage segment one by one. Finally, locate the voltage segment to estimate all parameters together based on preliminary estimations. The sketch of the SSO algorithm is shown in Fig.~\ref{fig:Algorithm}.

\section{Experimental Results}\label{sec:result}
Considering the wide use of ternary LIBs in transportation systems, a typical NCM811 (i.e., active particles in the positive electrode are made of \ce{LiNi_{0.8}Mn_{0.1}Co_{0.1}O_2} and those in the negative electrode are made of \ce{LiC_6}) 18650 battery cell is studied to evaluate the performance of parameter identification. The fresh cell has a capacity of 2200 mAh and is alternately discharged and charged under the 1 C-rate galvanostatic current for 2000 full cycles. 

The identification is conducted at the 0/500/1000/1500/2000-th cycles to validate the adaptability of this work under different SOH. Note that many parameters to be identified are immeasurable for real batteries. For ease of evaluation, the experiment is conducted on the AutoLion platform, a state-of-the-art commercial software widely used in academia and industry for lithium-ion battery studies~\cite{AutoLion}. Parameters of the battery are set to default values stored in the software for the NCM811 battery, which can be obtained in ref.~\cite{GU2023126192}. The simulation results are assumed as the battery measurements in this study. For the quasi-static test, the measuring interval is set to 200 seconds since the voltage change is almost unnoticeable under the 0.01 C-rate current. For the dynamic test, the measuring interval is set to around 0.5 seconds to track the volatile change of the voltage under the active status. After the measurements are obtained, the identification approach can be conducted without further processing. The identification approach is conducted on the MATLAB R2021 A platform in this work. The hardware for computation is a 2.11 GHz Intel Core i5-10210U processor with 16 GB of RAM.

Since the parameters to be identified are all scalars, we use the mean absolute error (MAE) as the accuracy metric as follows: MAE=$|\hat{\theta}-\theta|/\theta$.

\subsection{Quasi-static test result}
\begin{figure}[!htb]
    \centering
    \begin{minipage}{.24\textwidth}
        \centering
        \includegraphics[width=\textwidth]{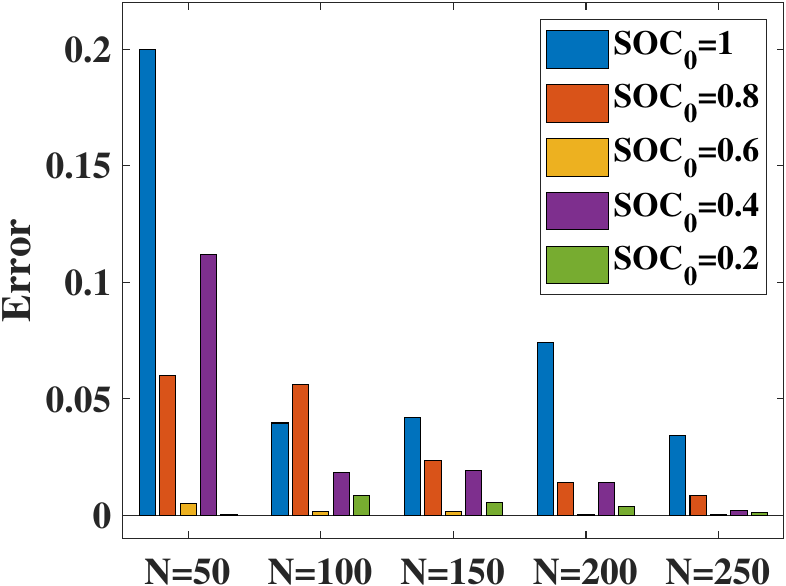}
        \subcaption{Fresh battery. }\label{fig:staticAccuracy_fresh}
    \end{minipage}%
    \begin{minipage}{.24\textwidth}
        \centering
        \includegraphics[width=\textwidth]{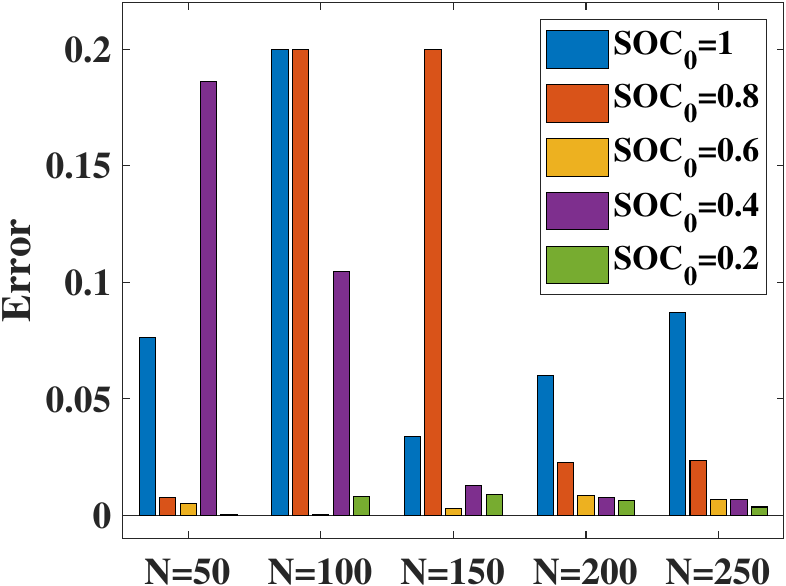}
        \subcaption{Aged battery (2000 cycles). }\label{fig:staticAccuracy_2000cycle}
    \end{minipage}
    \caption{Identification error of $\varepsilon_s^-$ in the quasi-static test with different settings}\label{fig:staticAccuracy_try}
\end{figure}

To shorten the experiment duration, a specific SOC range should be determined first to conduct the test. We conduct the tests at different SOC starting points with different durations. The identification errors of $\varepsilon_s^-$ at the new state and the aged state (2000 cycles) are plotted in Fig.~\ref{fig:staticAccuracy_try}.

It is observed that starting at the lower SOC with longer durations generally leads to higher accuracy. This finding is reasonable because the gradient of OCV is higher at lower SOCs, which makes the slight change in the voltage easier to be observed. Specifically, we find that starting around SOC$=0.6$ with the duration of $20000$s (i.e., the number of measurement points equals $100$) is relatively more accurate than other settings. It is noteworthy that the \ce{Li^+} stoichiometry range of active materials is unknown (i.e., $c_{s,\mathrm{max}}^{\pm}$ and $c_{s,\mathrm{min}}^{\pm}$ in Eq.(\ref{equ:thetalimit})), the SOC cannot be defined accurately. Thus, the OCV is used as an indicator to replace the SOC, i.e., the quasi-static test is recommended to start at OCV$=3.8$V (roughly near SOC$=0.6$) and lasts for $20000$s until $100$ measurement points are obtained.

To determine the proper solver for Eq.(\ref{equ:staticOpt}), both the exact algorithm (active set) and heuristic algorithms (PSO and GA) are tested. The hyper-parameters of the three algorithms are determined manually; typically, only the setting with the best performance is selected. For the active set algorithm, the optimality tolerance is set to 1e-6, the function tolerance is set to 1e-6, and the max function evaluation is set to 10000. For the PSO, the function tolerance is set to 1e-6, and the swarm size is set to 200. For the GA, the function tolerance is set to 1e-6, the max stall generation is set to 40, and the population size is set to 200. The identification result for $\varepsilon_s^-$ is presented in Table.~\ref{table:staticTest11}. Generally, the time cost of the Active Set method is significantly lower than heuristic algorithms. However, it is likely to be trapped in a local minimum. The heuristic algorithms perform better in searching for the global optimum. In addition, we find that the PSO performs better than the GA. The PSO is selected because of the trade-off between computation complexity and accuracy.
\begin{table}[!t]
	\caption{The comparison between different solvers on identifying $\varepsilon_s^-$.}\label{table:staticTest11}
	\resizebox{0.45\textwidth}{!}{\begin{tabular}{ccccc}
			\hline
			cycles & metric  & PSO & GA & Active Set \\ \hline
			\multirow{2}{*}{0} & Run time/s & 34.49 & 61.67  & \textbf{1.10} \\
			& MAE & \textbf{0.142\%} & 3.91\% & 19.38\%  \\
			\hline
			\multirow{2}{*}{500} & Run time/s & \textbf{3.25} & 18.41 & 26.24  \\
			& MAE    & \textbf{0.121\%}  & 4.35\% & 20.00\%    \\ \hline 
			\multirow{2}{*}{1000} & Run time/s & 3.18 & 18.01  & \textbf{0.91}       \\
			& MAE & \textbf{0.332\%}  & 1.44\% & 19.62\% \\ \hline 
			\multirow{2}{*}{1500} & Run time/s & 2.74   & 18.63  & \textbf{0.88}       \\
			& MAE    & \textbf{0.203\%}   & 1.04\% & 19.30\%   \\ \hline 
			\multirow{2}{*}{2000} & Run time/s & 3.64 & 16.44  & \textbf{1.26}       \\
			& MAE    & \textbf{0.058\%} & 2.10\% & 16.64\%      \\ \hline 
	\end{tabular}}
\end{table}

For the adopted PSO solver, the swarm size is set to $200$, the function tolerance is set to $10^{-6}$, and the lower and upper boundaries of the optimized variables are $80\%$ and $120\%$, respectively, of the actual values. The experimental results indicate that the average optimization time takes a few seconds. The identification accuracy of cells at different life stages is given in Table.~\ref{table:staticTest1}. For every parameter, the first row in bold font is the identified value, the second row is the actual value, and the third row is the error metrics.
\begin{table}[!t]
    \caption{The identification results of $\theta_s^{\pm,0}$ and $\varepsilon_s^{\pm}$.}\label{table:staticTest1}
    \resizebox{0.49\textwidth}{!}{\begin{tabular}{cccccc}
    \hline
    cycles             & 0              & 500            & 1000           & 1500           & 2000           \\ \hline
    \multirow{3}{*}{$\theta_s^{-,0}$} & \textbf{0.487} & \textbf{0.483} & \textbf{0.480} & \textbf{0.478} & \textbf{0.477} \\
                       & 0.486          & 0.482          & 0.480          & 0.478          & 0.477          \\
                       &0.099\%	&0.097\% &0.088\%	&0.089\%	&0.089\% \\
                       \hline
    \multirow{3}{*}{$\theta_s^{+,0}$} & \textbf{0.536} & \textbf{0.535} & \textbf{0.534} & \textbf{0.533} & \textbf{0.532} \\
                       & 0.536          & 0.535          & 0.534          & 0.533          & 0.532          \\ 
                       &0.023\%	&0.005\% &0.001\%	&0.008\%	&0.015\% \\\hline 
    \multirow{3}{*}{$\varepsilon_s^-$} & \textbf{0.666} & \textbf{0.654} & \textbf{0.643} & \textbf{0.633} & \textbf{0.624} \\
                       & 0.665          & 0.655          & 0.645          & 0.635          & 0.624          \\ 
                       &0.142\%	&0.121\% &0.332\%	&0.203\%	&0.089\% \\\hline 
    \multirow{3}{*}{$\varepsilon_s^+$} & \textbf{0.520} & \textbf{0.517} & \textbf{0.516} & \textbf{0.515} & \textbf{0.514} \\
                       & 0.519          & 0.516          & 0.515          & 0.514          & 0.513  \\ 
                       &0.144\%	&0.173\% &0.124\%	&0.103\%	&0.076\% \\
    \hline 
    \end{tabular}}
\end{table}
It is noteworthy that the initial \ce{Li^+} concentrations $c_s^{\pm,0}$ are replaced by the \ce{Li^+} stoichiometry $\theta_s^{\pm,0}$. They are equivalent, i.e., $c_s^{\pm,0}$=$\theta_s^{\pm,0}$$c_{s,\mathrm{max}}^{\pm}$, here $c_{s,\mathrm{max}}^{\pm}$ are determined by the chemical formulas of active particles~\cite{GU2023126192}.

\begin{figure}[!htb]
    \centering
    \begin{minipage}{.24\textwidth}
        \centering
        \includegraphics[width=\textwidth]{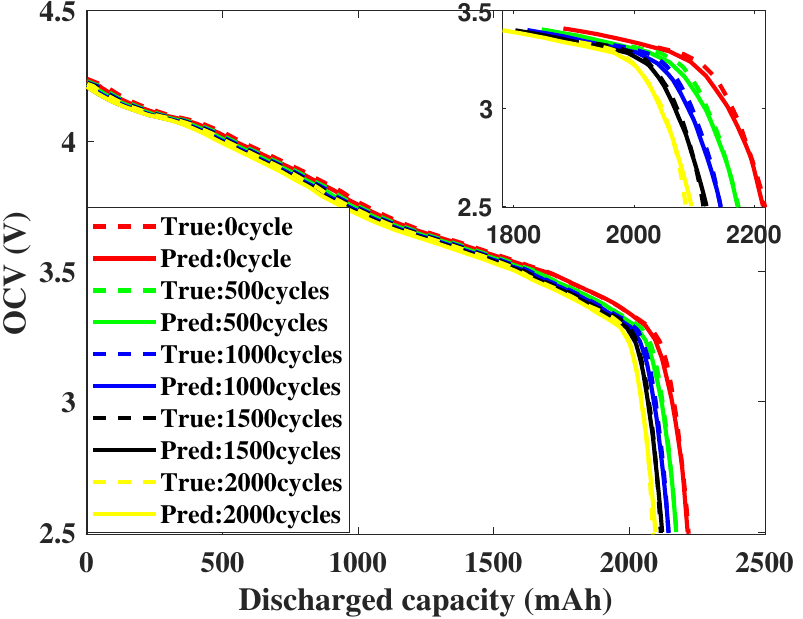}
        \subcaption{SOC-OCV curve. }\label{fig:SOC_OCV_result}
    \end{minipage}%
    \begin{minipage}{.24\textwidth}
        \centering
        \includegraphics[width=\textwidth]{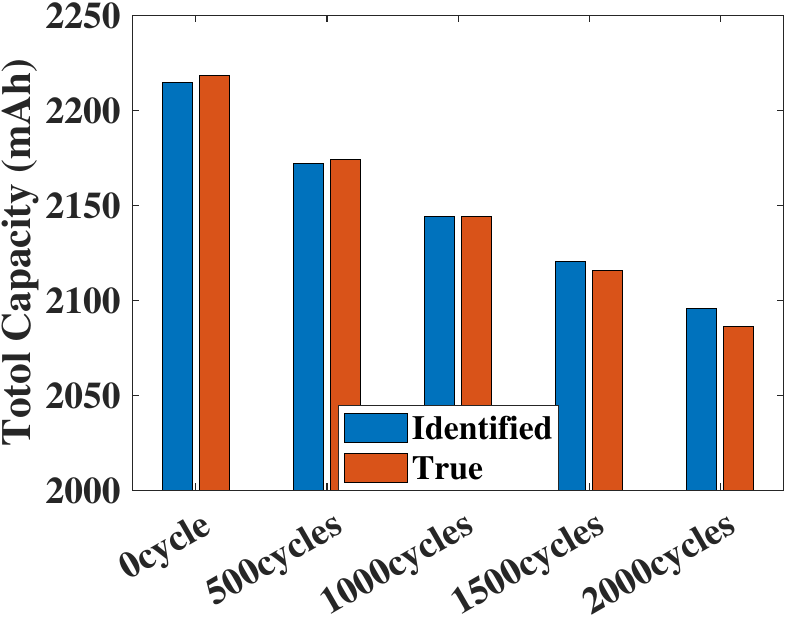}
        \subcaption{Total capacity. }\label{fig:Cap_result}
    \end{minipage}
    \caption{The identification result of macro parameters.}\label{fig:Cap_SOC_OCV_result}
\end{figure}

It is observed that all four parameters can be accurately and efficiently identified. Based on them, $C_Q$ and OCV curve can be derived, as shown in Fig.~\ref{fig:Cap_SOC_OCV_result}. Although the identification error increases gradually with the battery degradation, its absolute value generally meets the requirement, i.e., smaller than 0.5\%.

The remaining parameters, $\varepsilon_e^-$ and $R_f^{\pm}$, are identified via the empirical model formulated by Eqs. (\ref{equ:electrode_aged})-(\ref{equ:C6epsilone}). Coefficients of the empirical model are fitted on the simulation data from the AutoLion and listed in Table.~\ref{table:staticTest2}. 
\begin{table}[!t]
	\caption{The fitted coefficients of the empirical model.}\label{table:staticTest2}
    \resizebox{0.49\textwidth}{!}{\begin{tabular}{ccccccc}
    \hline
    \multirow{2}{*}{\begin{tabular}[c]{@{}c@{}}Fitted\\Coefficients\end{tabular}} & \begin{tabular}[c]{@{}c@{}}$k_f^+$\\(nm)\end{tabular} & \begin{tabular}[c]{@{}c@{}}$k_f^-$\\(nm)\end{tabular} & $k_e^-$ & $b_e^-$ & \begin{tabular}[c]{@{}c@{}}$\sigma_{f}^+$\\(S/m)\end{tabular}
     & \begin{tabular}[c]{@{}c@{}}$\sigma_{f}^-$\\(S/m)\end{tabular} \\
     & -7.68e3 & -3.76e3 & 6.00 & 0.659 & 1.52e-5 & 1.54e-5 \\ 
    \hline
    \end{tabular}}
\end{table}
\begin{table}[!t]
	\caption{The identification results of $\varepsilon_e^{-}$ and $R_f^{\pm}$.}\label{table:staticTest3}
	\resizebox{0.49\textwidth}{!}{\begin{tabular}{cccccc}
			\hline
			cycles   & 0    & 500  & 1000 & 1500  & 2000  \\ \hline
			\multirow{3}{*}{$\varepsilon_e^-$} & \textbf{0.2833} & \textbf{0.2763} & \textbf{0.2710} & \textbf{0.2679} & \textbf{0.2658} \\
			& 0.2827 & 0.2762 & 0.2720  & 0.2686  & 0.2656          \\
			&0.22\%	&0.02\% &0.36\%	&0.26\%	&0.04\% \\
			\hline
			\multirow{3}{*}{\begin{tabular}[c]{@{}c@{}}$R_f^-$\\($10^{-4}\Omega$m$^2$)\end{tabular}} & \textbf{0} & \textbf{17.32} & \textbf{30.42} & \textbf{38.25} & \textbf{43.49} \\
			& 0 & 19.26 & 29.80 & 38.28 & 45.56          \\ 
			& / &10.1\% &2.07\%	&0.09\%	&4.56\% 
			\\\hline 
			\multirow{3}{*}{\begin{tabular}[c]{@{}c@{}}$R_f^+$\\($10^{-4}\Omega$m$^2$)\end{tabular}} & \textbf{0} & \textbf{13.24} & \textbf{20.99} & \textbf{26.53} & \textbf{31.22} \\
			& 0 & 15.45 & 21.86 & 26.78 & 30.93 
			 \\ 
			& / &14.3\% &3.96\%	&0.94\%	& 0.95\% \\
			\hline 
	\end{tabular}}
\end{table}

The identification result is presented in Table.~\ref{table:staticTest3}. It can be observed that $\varepsilon_{e}^-$ can be identified accurately for the battery under different life stages. Since the initial thickness of the SEI/CEI film is small, the identification error of $R_f^{\pm}$ is relatively high in the early stage of the whole lifespan. However, with the growth of the SEI/CEI film during the battery degradation, the estimation accuracy gradually increases, indicating the effectiveness of the empirical model.
As an important indicator of the battery SOH, it is more meaningful to identify $R_f^{\pm}$ for aged batteries, and a relatively large error in the early life stage is acceptable. However, the main limitation of the empirical model should also be stated. The empirical model relies on the fitted coefficients, which are assumed to be available from the manufacturer or the material experiments. When these values are unknown, empirical values must be used, and thus, the increment of the identification error is unavoidable.

\subsection{Sensitivity analysis}
To obtain a reliable result, the sampling ranges of parameters identified in the dynamic test are broad, as shown in Table.~\ref{table:samplingrange}. Note also that the ranges are asymmetric around the actual values because, during the battery degradation, the conductive or diffusion parameters gradually decrease while the ohmic parameters increase progressively. Thus, the range is extended in the direction of the parameter-changing trajectory in degradation.
\begin{table}[!t]
	\caption{The sampling ranges of parameters identified in the dynamic test.}\label{table:samplingrange}
	\resizebox{0.49\textwidth}{!}{
	\begin{tabular}{cccccc}
		\hline
		params       & true value     & sampling range         & params       & true value & sampling range          \\ \hline
		$R_c$        & 6.4e-3 & $\left[0,0.05\right]$  & $\sigma_s^-$ & 66.5       & $\left[6.6,100\right]$  \\
		$D_s^+$      & 1              & $\left[0.1,1.5\right]$ & $t_+^0$      & 0.38       & $\left[0.2,0.45\right]$ \\
		$\sigma_s^+$ & 1.97           & $\left[0.2,3\right]$   & $\kappa$     & 1          & $\left[0.1,1.5\right]$  \\
		$D_s^-$      & 1              & $\left[0.1,1.5\right]$ & $D_e$        & 1          & $\left[0.1,1.5\right]$  \\ \hline
	\end{tabular}}
\end{table}

\begin{figure}[!htb]
	\centering
\begin{minipage}{.24\textwidth}
		\centering
		\includegraphics[width=\textwidth]{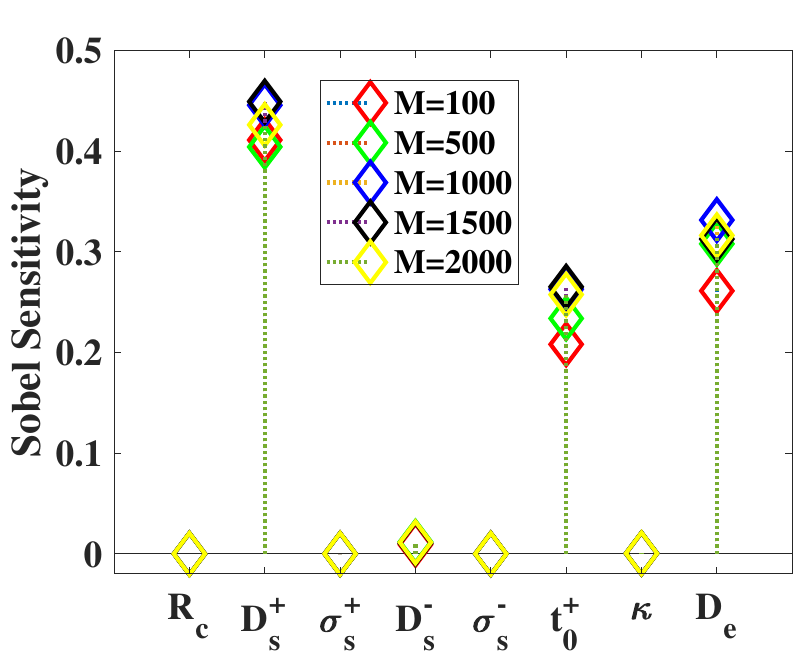}
\subcaption{Sampling scale. }\label{fig:SobolSamplingScale_excitation}
\end{minipage}%
\begin{minipage}{.25\textwidth}
		\centering
		\includegraphics[width=\textwidth]{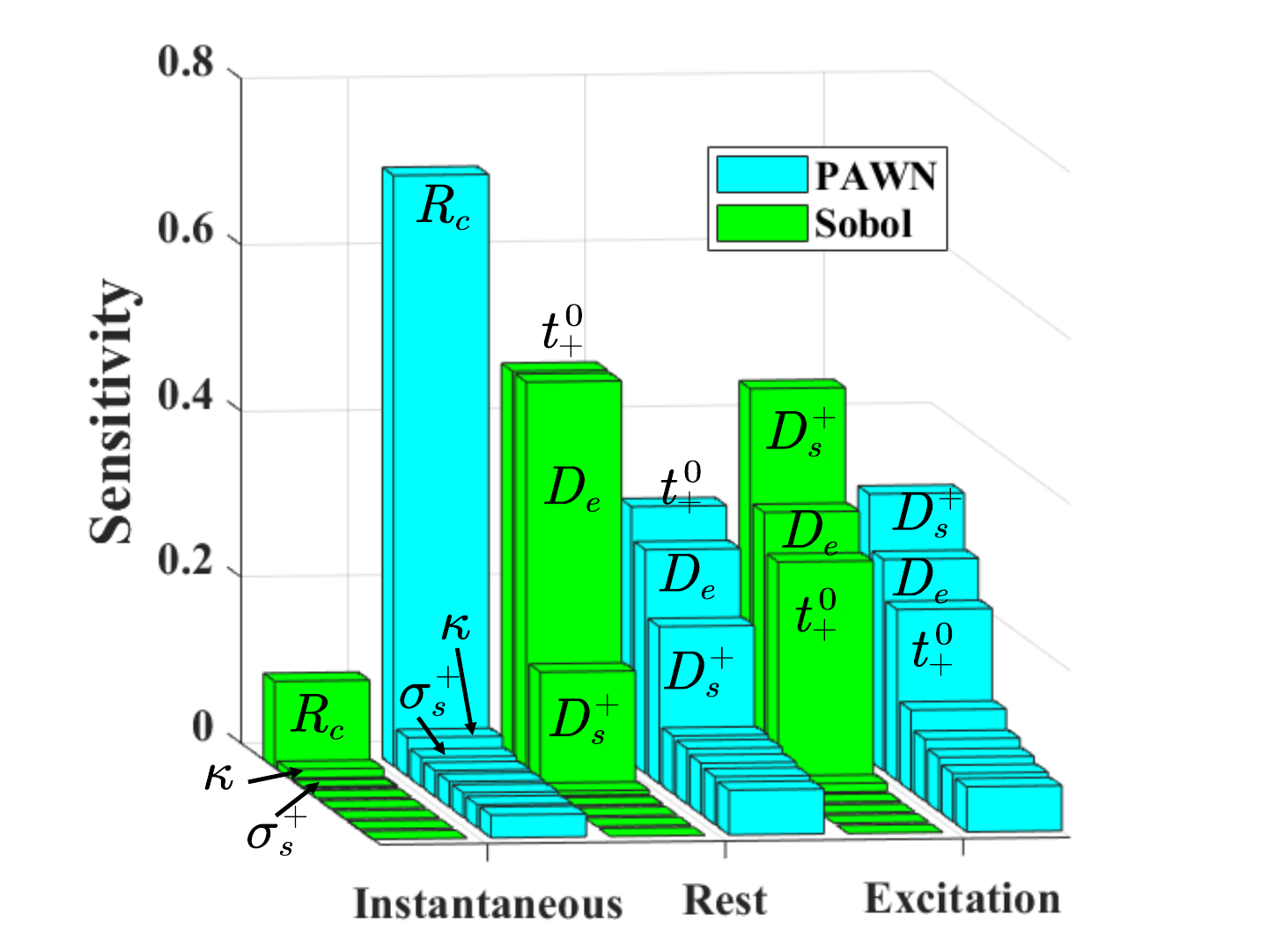}
\subcaption{Sensitivity computation method. }\label{fig:SobolSamplingScale_relaxation}
\end{minipage}
	\caption{Sensitivity analysis experiment.}\label{fig:Sobolsamplingscale}
\end{figure}

First, the proper sampling scale $M$ should be determined. Take the pulse with a duration of 220 seconds as an example, the sensitivity of parameters to the voltage segment in the excitation process when $M$=100, 500, 1000, 1500, 2000 is plotted in the left of Fig.~\ref{fig:Sobolsamplingscale}.
It is observed that when $M$ exceeds 1000, the Sobol sensitivity tends to reach convergence. To realize a trade-off between reliability and cost, $M$ is set to 1000 in this work. In contrast to a prevalent density-based method, PAWN~\cite{PIANOSI20151}, the Sobol method also performs well, as shown in the right of Fig.~\ref{fig:Sobolsamplingscale}. The first three parameters of sensitivity obtained by the two methods on different voltage segments are the same. Generally, the Sobol method is easier to distinguish between various parameters, e.g., in the instantaneous process, $R_c$ and $\kappa$ can be explicitly distinguished from the other parameters. Meanwhile, the Sobol method directly uses the objective function in Eq.(\ref{equ:dynamicOpt}) as the input, which ensures the consistency between sensitivity analysis and the optimization approach. Thus, the Sobol method is adopted to compute the sensitivity. The result is shown in Table.~\ref{table:Sobolresult}.
\begin{table*}[!htb]
	\centering
	\caption{The Sobol sensitivity of parameters to different voltage segments when $M$=1000.}\label{table:Sobolresult}
	\resizebox{0.98\textwidth}{!}{\begin{tabular}{ccccccccc}
			\hline
			& $R_c$                            & \cellcolor[HTML]{FFFFFF}$D_s^+$  & $\sigma_s^+$ & $D_s^-$  & $\sigma_s^-$ & $t_0^+$                          & $\kappa$                         & $D_e$                            \\ \hline
			$\zeta_1$                         & \cellcolor[HTML]{34FF34}1.04E-01 & 5.98E-06                         & \cellcolor[HTML]{C0C0C0}1.93E-03     & \cellcolor[HTML]{C0C0C0}1.67E-08 & \cellcolor[HTML]{C0C0C0}2.52E-06     & 4.71E-05                         & 1.06E-02                         & 4.81E-06                         \\
			\cellcolor[HTML]{FD6864}$\zeta_2$ & 1.03E-01                         & 1.09E-05                         & \cellcolor[HTML]{C0C0C0}1.92E-03     & \cellcolor[HTML]{C0C0C0}7.14E-08 & \cellcolor[HTML]{C0C0C0}2.53E-06     & 9.01E-05                         & \cellcolor[HTML]{34FF34}1.06E-02 & 1.56E-07                         \\
			$\zeta_3$                         & 1.02E-01                         & 1.16E-05                         & \cellcolor[HTML]{C0C0C0}1.93E-03     & \cellcolor[HTML]{C0C0C0}7.76E-07 & \cellcolor[HTML]{C0C0C0}2.44E-06     & 1.56E-04                         & 1.05E-02                         & 1.89E-05                         \\
			$\zeta_4$                         & 1.01E-01                         & 1.60E-06                         & \cellcolor[HTML]{C0C0C0}1.81E-03     & \cellcolor[HTML]{C0C0C0}3.44E-06 & \cellcolor[HTML]{C0C0C0}2.49E-06     & 1.91E-04                         & 1.03E-02                         & 9.20E-05                         \\\hline
			$\zeta_5$                         & 3.86E-28                         & 4.82E-02                         & \cellcolor[HTML]{C0C0C0}4.85E-05     & \cellcolor[HTML]{C0C0C0}9.77E-05 & \cellcolor[HTML]{C0C0C0}3.76E-06     & \cellcolor[HTML]{34FF34}8.03E-01 & 5.03E-03                         & 1.30E-01                         \\
			$\zeta_6$                         & 8.46E-29                         & 1.19E-01                         & \cellcolor[HTML]{C0C0C0}2.42E-05     & \cellcolor[HTML]{C0C0C0}3.58E-04 & \cellcolor[HTML]{C0C0C0}1.59E-06     & 6.13E-01                         & 2.29E-03                         & 2.21E-01                         \\
			\cellcolor[HTML]{FD6864}$\zeta_7$ & 2.95E-29                         & 2.61E-01                         & \cellcolor[HTML]{C0C0C0}1.47E-05     & \cellcolor[HTML]{C0C0C0}1.52E-03 & \cellcolor[HTML]{C0C0C0}4.58E-07     & 3.99E-01                         & 7.58E-04                         & 3.09E-01                         \\
			$\zeta_8$                         & 1.01E-29                         & 4.00E-01                         & \cellcolor[HTML]{C0C0C0}6.04E-06     & \cellcolor[HTML]{C0C0C0}1.15E-02 & \cellcolor[HTML]{C0C0C0}9.26E-08     & 2.47E-01                         & 4.21E-04                         & 3.09E-01                         \\\hline
			$\zeta_9$                         & 0.00E+00                         & \cellcolor[HTML]{34FF34}6.74E-01 & \cellcolor[HTML]{C0C0C0}5.05E-05     & \cellcolor[HTML]{C0C0C0}4.54E-03 & \cellcolor[HTML]{C0C0C0}7.76E-06     & 7.82E-01                         & 9.65E-03                         & 8.18E-02                         \\
			$\zeta_{10}$                      & 0.00E+00                         & 4.15E-01                         & \cellcolor[HTML]{C0C0C0}2.40E-05     & \cellcolor[HTML]{C0C0C0}4.40E-03 & \cellcolor[HTML]{C0C0C0}4.20E-06     & 6.50E-01                         & 6.55E-03                         & 2.19E-01                         \\
			$\zeta_{11}$                      & 0.00E+00                         & 1.90E-01                         & \cellcolor[HTML]{C0C0C0}2.44E-05     & \cellcolor[HTML]{C0C0C0}2.38E-03 & \cellcolor[HTML]{C0C0C0}9.40E-07     & 4.92E-01                         & 1.34E-03                         & 3.41E-01                         \\
			$\zeta_{12}$                      & 0.00E+00                         & 1.24E-01                         & \cellcolor[HTML]{C0C0C0}5.82E-06     & \cellcolor[HTML]{C0C0C0}5.97E-03 & \cellcolor[HTML]{C0C0C0}1.94E-07     & 4.40E-01                         & 5.70E-04                         & \cellcolor[HTML]{34FF34}4.64E-01 \\\hline
	\end{tabular}}
\end{table*}

Based on the sensitivity analysis result, the parameters identified in this stage can be divided into three groups. The first group contains $\sigma_s^{\pm}$ and $D_s^-$, which have little impact on all voltage segments (the Sobol sensitivity is smaller than 0.01). The low sensitivity of $\sigma_s^{\pm}$ is because the active particles are commonly mixed up with conductive filler additives and can be viewed as good conductors, inducing the ohmic potential drop in the solid-phase negligible compared to other components in the voltage. The low sensitivity of $D_s^-$ is attributed to the slight gradient of $U_{\mathrm{OCP}}^-$ (as shown in Fig.~\ref{fig:staticV}). Specifically, $D_s^-$ controls the diffusion process of Li-ions in active particles and determines Li-ion concentrations on the surface of active particles, denoted by $c_{ss}^-$, which appears explicitly in the voltage as a term of $U_{\mathrm{OCP}}^-\left(c_{ss}^-\right)$~\cite{GU2023126192}. When the gradient of $U_{\mathrm{OCP}}^-$ is minimal, the change in $c_{ss}^-$ can hardly be observed from the voltage. With such low sensitivity, the identification results are likely to be unreasonable. Thus, $\sigma_s^{\pm}$ and $D_s^-$ are removed from the identification task and set to empirical values.

The second group, i.e., $\boldsymbol{\Theta}^{\mathrm{I}}$, contains $R_c$ and $\kappa$. They are sensitive to the voltage segments in the instantaneous process. It is reasonable because the transport of electrons is fast and can be quickly reflected in the voltage when the current suddenly changes. The third group, i.e., $\boldsymbol{\Theta}^{\mathrm{T}}$, contains $D_s^+$, $t_0^+$ and $D_e$. They are sensitive to the transition process. This is because $D_s^+$ dominates the diffusion process of \ce{Li^+} in active particles of the positive electrode, $t_0^+$ and $D_e$ dominate the migration process of \ce{Li^+} in the electrolyte. Compared to electrons, the transport speed of Li-ions is much slower, their impacts on the voltage are gradually reflected in the voltage, exhibiting a transition process. To conclude, we find that the sensitivity analysis result is consistent with the physical interpretations, which makes the analysis result more reliable. 

\subsection{Dynamic test result}
For clarity, in Table.~\ref{table:Sobolresult}, each parameter along with the corresponding voltage segment where it has the highest sensitivity is painted green. The voltage segments where parameters in $\boldsymbol{\Theta}^{\mathrm{I}}$ and $\boldsymbol{\Theta}^{\mathrm{T}}$ have the smallest sensitivity differences are painted red. According to the SSO algorithm, the solving process contains seven steps in all:

1) the preliminary estimation of $R_c$ is conducted on $\zeta_1$, 

2) the preliminary estimation of $\kappa$ is conducted on $\zeta_2$, 

3), final values of $R_c$ and $\kappa$ are identified on $\zeta_2$, 

4), the preliminary estimation of $t_0^+$ is conducted on $\zeta_5$, 

5), the preliminary estimation of $D_s^+$ is conducted on $\zeta_9$, 

6), the preliminary estimation of $D_e$ is conducted on $\zeta_{12}$, 

7), final values of $D_s^+$, $t_0^+$ and $D_e$ are identified on $\zeta_7$. 

\begin{figure*}[!htb]
	\centering
\begin{minipage}{.8\textwidth}
		\centering
		\includegraphics[width=\textwidth]{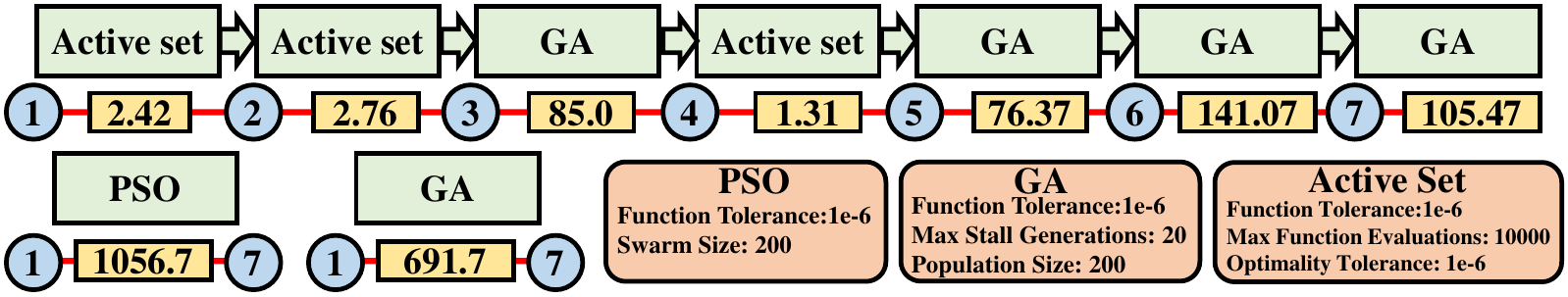}
\end{minipage}%
\caption{The solver selection and the computation time in each step.}\label{fig:SSHO}
\end{figure*}

To select the proper solver for each identification step, we compared the performance of the exact algorithm and heuristic algorithms. The result is shown in Table.~\ref{table:dynamicTest11}.
\begin{table}[!t]
	\caption{The comparison between different solvers in every identifictaion step on a fresh battery.}\label{table:dynamicTest11}
	\resizebox{0.45\textwidth}{!}{\begin{tabular}{ccccc}
			\hline
			Step               & Metric        & PSO     & GA               & Active Set       \\ \hline
			\multirow{2}{*}{1} & Run time/s    & 146.441 & 77.722           & \textbf{2.707}   \\
			& MAE           & 53.07\% & \textbf{52.47\%} & 53.04\%          \\ \hline
			\multirow{2}{*}{2} & Run time/s    & 92.93
			   & 	92.12            & \textbf{1.592}   \\
			& MAE           & 17.47\% & 17.47\%          & \textbf{17.47\%} \\ \hline
			\multirow{2}{*}{3} & Run time/s    & 102.873 & \textbf{82.622}  & 114.111          \\
			& MAE($\kappa$) & 0.35\%  & \textbf{0.35\%}  & 0.35\%           \\ \hline
			\multirow{2}{*}{4} & Run time/s    & 77.863  & 72.455           & \textbf{1.034}   \\
			& MAE           & 0.65\%  & \textbf{0.64\%}  & 0.65\%           \\ \hline
			\multirow{2}{*}{5} & Run time/s    & 82.996  & 74.086           & \textbf{0.295}   \\
			& MAE           & 30.68\% & \textbf{28.21\%} & 35.00\%          \\ \hline
			\multirow{2}{*}{6} & Run time/s    & 149.126 & 140.562          & \textbf{1.9}     \\
			& MAE           & 30.08\% & \textbf{30.08\%} & 35.00\%          \\ \hline
			\multirow{2}{*}{7} & Run time/s    & 136.148 & \textbf{102.075} & 148.955          \\
			& MAE($t_0^+$)  & 5.79\%  & \textbf{1.84\%}  & 16.04\%      \\
			 \hline   
	\end{tabular}}
\end{table}
In steps (1) and (2), the active set is much faster than the heuristic algorithms, and its accuracy is slightly lower. Since it is a preliminary estimation, a larger error is tolerated. The active set algorithm is selected in this step. In step (3), $R_c$ and $\kappa$ need to be identified. The GA performs best in the run time and is selected. In step (4), the situation is similar to that in step (1); the active set is selected. In steps (5) and (6), the active set fails to search for the global optimum. Thus, the GA is selected. In step (7), the GA performs best in the speed and accuracy; thus, it is selected. The selection of the proper solver and the specific setting for each step is shown in Fig.~\ref{fig:SSHO}. To more clearly demonstrate the efficiency improvement of the SSO algorithm, the comparison between the pure PSO and GA is presented in Fig.~\ref{fig:SSHO}. If the GA or PSO is applied in every step throughout the identification, the entire time cost is 1056.7 seconds or 691.7 seconds, respectively. The SSO algorithm reduces time as the whole cost to 414.4 seconds (the computation time in each step is given in the box between adjacent nodes). It is also noteworthy that the pure active set is not presented here since it can hardly find the optimum at steps (3) and (7).

The identification result for the battery at different life stages is presented in Table.~\ref{table:dynamicTest1}, where the first row of each parameter refers to the identified value and the second row refers to the MAE ( the actual values are listed in Table.~\ref{table:samplingrange}). It is observed that the MAEs of $R_c$, $D_s^+$ and $D_e$ are relatively large. This originates in the inherent error of the simplified model in ref.~\cite{GU2023126192}. Specifically, the identification results of $R_c$, $D_s^+$ and $D_e$ are impacted by the reaction rate distribution approximation, solid-phase diffusion approximation, and solution-phase diffusion approximation, respectively. Since the identification aims to minimize the voltage error, it can introduce bias to these parameters. As shown in Table.~\ref{table:dynamicTest2}, we can see that the voltage prediction errors are at least reduced by half compared with the actual parameters. Thus, the biases in the identified values of $R_c$, $D_s^+$ and $D_e$ are reasonable and necessary; they can be viewed as a patch of the original simplified model. In addition, it is observed that the bias of $R_c$ decreases roughly with battery degradation, inspiring us to treat it as an indicator of the SOH.
\begin{table}[!t]
	\caption{The identification result of the dynamic test.}\label{table:dynamicTest1}
	\resizebox{0.47\textwidth}{!}{\begin{tabular}{cccccc}
			\hline
			cycles   & 0    & 500  & 1000 & 1500  & 2000  \\ \hline
			\multirow{2}{*}{\begin{tabular}[c]{@{}c@{}}$R_c$\\($10^{-3}\Omega$m$^2$)\end{tabular}} 
			&8.58 & 8.26 & 7.45 & 6.88 & 6.91 \\
			&\textbf{34.1\%} &\textbf{29.1\%}&\textbf{16.4\%}&\textbf{7.50\%}&\textbf{7.97\%} \\	\hline
			\multirow{2}{*}{$\kappa$} 
			&0.997 & 1.041 & 1.052 & 0.988 & 1.002 \\
			&\textbf{0.3\%} &\textbf{4.1\%}&\textbf{5.2\%}&\textbf{1.2\%}&\textbf{0.2\%}  \\ \hline 
			\multirow{2}{*}{$D_s^+$} 
			&1.416 & 1.384 & 1.376 & 1.386 & 1.390 \\
			&\textbf{41.6\%} &\textbf{38.4\%}&\textbf{37.6\%}&\textbf{38.6\%}&\textbf{39.0\%}  \\ \hline 
			\multirow{2}{*}{$t_0^+$} 
			&0.373 & 0.384 & 0.370 & 0.365 & 0.368 \\
			&\textbf{1.84\%} &\textbf{1.05\%}&\textbf{2.63\%}&\textbf{3.95\%}&\textbf{3.16\%}  \\ \hline 
			\multirow{2}{*}{$D_e$} 
			&1.296 & 1.254 & 1.283 & 1.283 & 1.273 \\
			& \textbf{29.6\%} &\textbf{25.4\%}&\textbf{28.3\%}&\textbf{28.3\%}&\textbf{27.3\%}  \\ \hline
			Run time(s) & 414 & 422 & 427 & 418 & 439 \\ \hline
	\end{tabular}}
\end{table}
\begin{table}[!t]
	\caption{The voltage RMSE ($\times$10$^{-3}$V) with identified parameters.}\label{table:dynamicTest2}
	\resizebox{0.45\textwidth}{!}{\begin{tabular}{cccccc}
			\hline
			cycles   & 0    & 500  & 1000 & 1500  & 2000  \\ \hline
			\multirow{2}{*}{Pulse 1 ($\zeta_1,\zeta_5,\zeta_9$)} 
			&2.4 & 3.9 & 4.8 & 5.7 & 6.5 \\
			&\textbf{0.72} &\textbf{0.66}&\textbf{0.66}&\textbf{0.70}&\textbf{0.69} \\	\hline
			\multirow{2}{*}{Pulse 2 ($\zeta_2,\zeta_6,\zeta_{10}$)} 
			&3.2 & 4.7 & 6.0 & 7.1 & 8.1 \\
			&\textbf{0.97} &\textbf{0.91}&\textbf{0.95}&\textbf{0.96}&\textbf{0.95}  \\ \hline 
			\multirow{2}{*}{Pulse 3 ($\zeta_3,\zeta_7,\zeta_{11}$)} 
			&3.9 & 5.0 & 6.4 & 7.7 & 9.0 \\
			&\textbf{1.41} &\textbf{1.42}&\textbf{1.39}&\textbf{1.43}&\textbf{1.41}  \\ \hline 
			\multirow{2}{*}{Pulse 4 ($\zeta_4,\zeta_8,\zeta_{12}$)} 
			& 4.1 & 4.2 & 5.8 & 7.5 & 9.1 \\
			&\textbf{2.04} &\textbf{2.17}&\textbf{2.08}&\textbf{2.09}&\textbf{2.03}  \\ \hline 
	\end{tabular}}
\end{table}

\begin{figure}[!htb]
	\centering
\begin{minipage}{.24\textwidth}
		\centering
		\includegraphics[width=\textwidth]{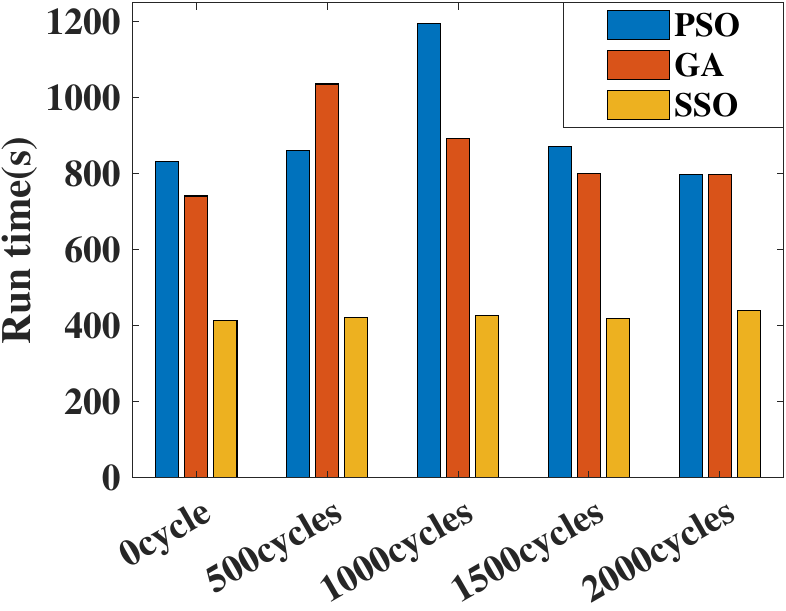}
\subcaption{Run time. }\label{fig:compare_time}
\end{minipage}%
\begin{minipage}{.24\textwidth}
		\centering
		\includegraphics[width=\textwidth]{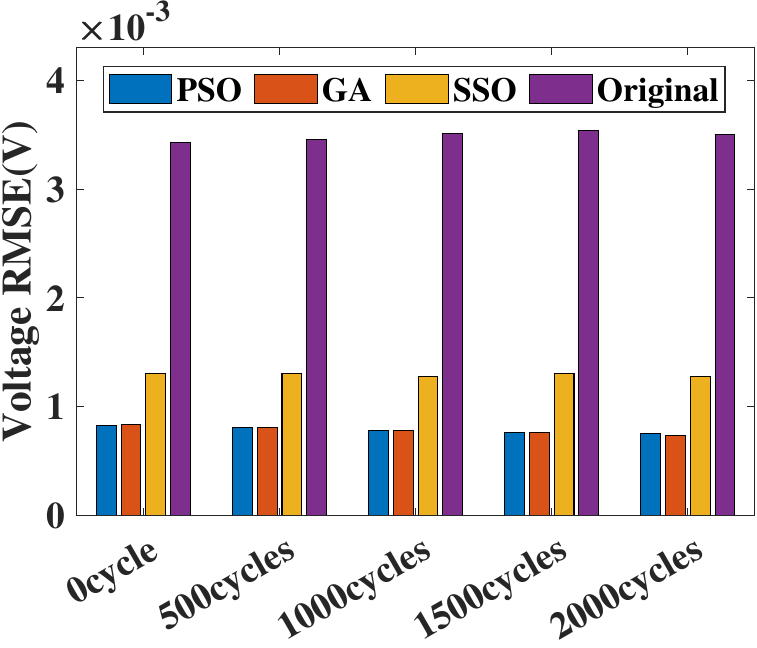}
\subcaption{Voltage RMSE}\label{fig:compare_V}
\end{minipage}
\caption{Comparison between SSO, PSO and GA.}\label{fig:Compare}
\end{figure}

The SSO algorithm is also compared to directly applying the classic PSO and GA, as shown in Fig.~\ref{fig:Compare}. It can be observed that the computation time can be reduced by approximately half to search for an acceptable solution, which can efficiently reduce the burden of practical implementation.

\subsection{Model validation}
\begin{figure}[!htb]
	\centering
	\begin{minipage}{.24\textwidth}
		\centering
		\includegraphics[width=\textwidth]{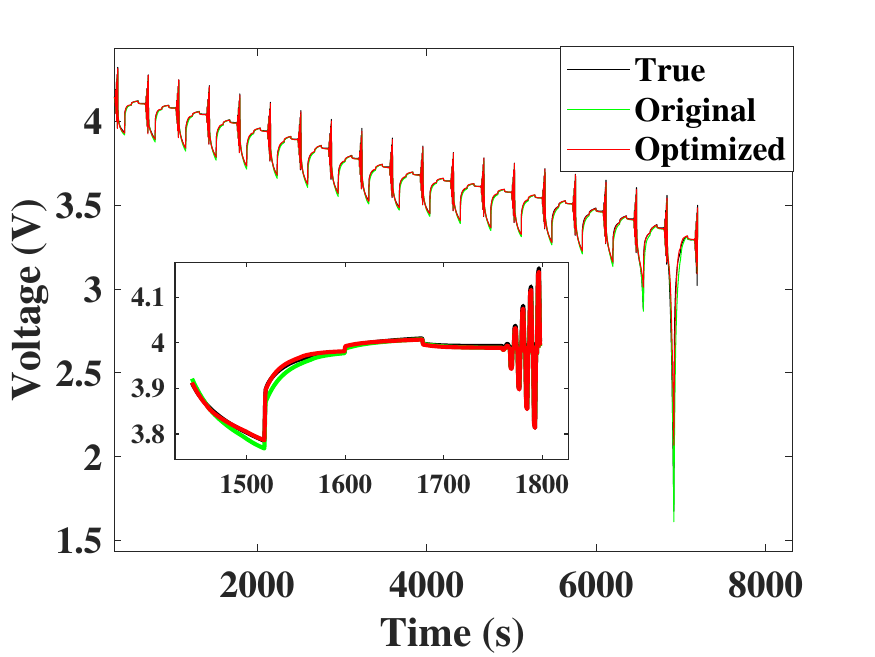}
		\subcaption{Pulsed discharge. }\label{fig:Vt_Discharge}
	\end{minipage}%
	\begin{minipage}{.22\textwidth}
		\centering
		\includegraphics[width=\textwidth]{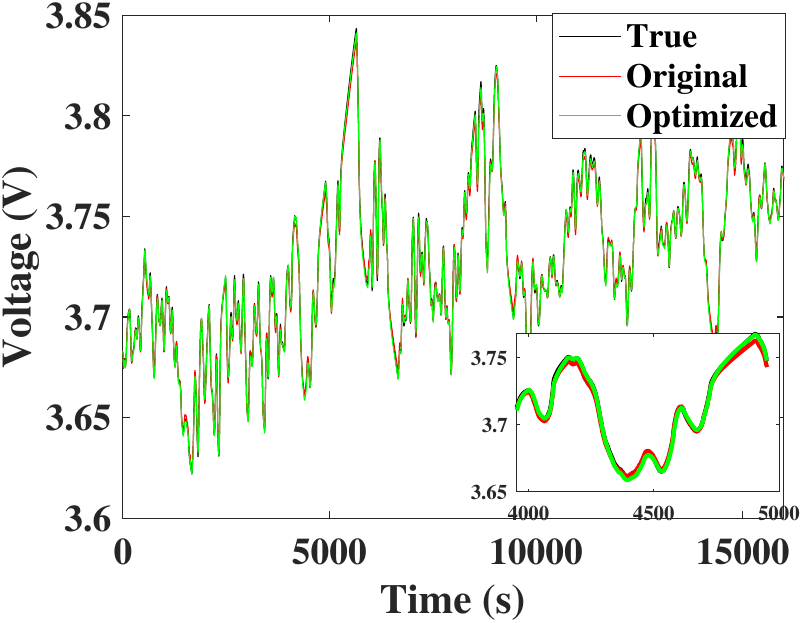}
		\subcaption{Regulation service. }\label{fig:Vt_Reg}
	\end{minipage}
	\caption{The predicted voltage of the optimized battery model in typical working protocols.}\label{fig:Vt}
\end{figure}

After all the parameters are identified, two typical working protocols are selected to validate the model using the identified parameters, as shown in Fig.~\ref{fig:Vt}. The first protocol is composed of pulsed discharge currents. The entire SOC range $[0,1]$ is divided into 20 parts evenly. Within each 5\% SOC range, the battery is first discharged by galvanostatic currents and then the pulsed currents. The galvanostatic current can simulate the battery under the normal use, such as driving with a uniform speed. In contrast, the pulsed current can simulate the battery under an emergency, such as the accelerating or braking. The second protocol refers to a typical scenario where the battery provides regulation service; the regulation signal is from the Pennsylvania-New Jersey-Maryland (PJM) electricity market. We choose the regulation scenario for two reasons. First, there have been industry applications using electric vehicles to connect to the grid and provide regulation services~\cite{peng2017dispatching}. Second, the regulation profile is highly random and appropriate to be used to evaluate the robustness of our method. The voltage profiles are plotted in Fig.~\ref{fig:Vt}. The RMSE of the simulated voltage and the actual voltage is calculated and listed in Table.~\ref{table:dynamicTest3}. It is observed that the model accuracy is improved after the parameterization in both scenarios.
\begin{table}[!t]
	\caption{The battery model accuracy improvement.}\label{table:dynamicTest3}
	\resizebox{0.48\textwidth}{!}{\begin{tabular}{ccccc}
			\hline
			& \multicolumn{2}{c}{Pulsed discharge} & \multicolumn{2}{c}{Regulation service} \\ \hline
			Params  & Original      & Identified           & Original       & Identified            \\ \hline
			RMSE(V) & 0.00264       & \textbf{0.00147}     & 0.03020        & \textbf{0.02380}      \\ \hline
	\end{tabular}}
\end{table}

To conclude, the proposed work can efficiently identify the electrochemical parameters of the LIB at different life stages. The test time and computation time are reduced considerably compared to existing methods. The identified parameters have concrete physical interpretations and can be used as the SOH indicator. However, there remain several drawbacks. First, the solver selection and hyper-parameter tuning are conducted manually, which might not be the best option. Second, the robustness and generality of the identification approach need further investigation. Third, some parameters with low sensitivity to the voltage are still unable to be observed. Additional measurements must be investigated to identify these parameters, e.g., the temperature, the geometric measurement, etc.

\section{Conclusion}\label{sec:con}
This paper proposes a parameter identification approach for the electrochemical LIB model, which considerably reduces the time cost of the test and the computation. The parameters are first grouped manually based on the physical properties and assigned to two sequenced tests for identification. The two tests, named the quasi-static test and the dynamic test, are compressed on time for practical implementation. Proper optimization models and an SSO algorithm are developed to search for the optimal parameters efficiently. Typically, the Sobol method is applied to give the sensitivity analysis result. Based on the sensitivity indexes, the SSO algorithm can decouple the mixed impacts of different parameters during the identification. For validation, numerical experiments on a typical NCM811 battery at different life stages are conducted. The proposed approach saves about half the time finding the proper parameter value. The identification accuracy of crucial parameters related to battery degradation can exceed 95\%. Case study results indicate that the identified parameters can not only improve the accuracy of the battery model but also be used as the indicator of the battery SOH. However, there remain some limitations of this work. First, identifying some parameters relies on the empirical model, whose generality and accuracy might not be ensured. Second, some parameters with low sensitivity to the voltage are still unable to be observed. In future work, we will focus on identifying them based on more measurement signals like temperature and pressure. Third, the solver selection in each step of the SSO algorithm is tuned manually, which lacks generality and robustness.

% or


%\appendices
%% you can choose not to have a title for an appendix
%% if you want by leaving the argument blank
%\section{}

%\section*{Acknowledgment}

\ifCLASSOPTIONcaptionsoff
  \newpage
\fi



\bibliographystyle{IEEEtran}
\bibliography{literature}
\end{document}